\documentclass[12pt]{article}
\usepackage{amsmath}
\usepackage{amssymb,amsfonts}
\usepackage[all]{xy}
\usepackage{graphicx}
\usepackage[parfill]{parskip}
\usepackage[usenames]{color}
\usepackage{sidecap}
\usepackage{pdfpages}
\usepackage{tikz}
\usepackage{comment}
\usetikzlibrary{shapes.geometric}
\usetikzlibrary{snakes}
\usepgflibrary{arrows}
\usepackage{enumitem}
\usepackage{mheck}

\DeclareFontFamily{U}{mathc}{}
\DeclareFontShape{U}{mathc}{m}{it}%
{<->s*[1.03] mathc10}{}

\DeclareMathAlphabet{\mathscr}{U}{mathc}{m}{it}

\numberwithin{equation}{section}
\numberwithin{table}{section}

\newcommand{\be}{\begin{equation}}
\newcommand{\ee}{\end{equation}}
\def\beq{\begin{eqnarray}}
\def\eeq{\end{eqnarray}}
\def\ba{\begin{eqnarray}}
\def\ea{\end{eqnarray}}

\def\ep1{\epsilon_1}
\def\eps2{\epsilon_2}

\newcommand{\IZ}{\mathbb{Z}}
\newcommand{\IC}{\mathbb{C}}
\newcommand{\IP}{\mathbb{P}}
\newcommand{\IN}{\mathbb{N}}
\newcommand{\IR}{\mathbb{R}}

\newcommand{\IF}{\mathbb{F}}

\newcommand{\Ztop}{\mathrm{Z_{\text{top}}}}

\newcommand{\nn}{\nonumber}

\newcommand{\pFf}{P^{F\rightarrow f}_{\bk}}
\newcommand{\pfZ}{P^{f\rightarrow Z}_{\bk}}
\newcommand{\pZf}{P^{Z\rightarrow f}_{\bk}}

\newcommand{\cO}{{\cal O}}

\newcommand{\cF}{{\cal F}}

\newcommand{\cZ}{{\cal Z}}

\newcommand{\lc}[2]{D_{#1,#2}}

\newcommand{\tb}{\mathscr{b}}
\newcommand{\tc}{\mathscr{c}}
\newcommand{\td}{\mathscr{d}}
\newcommand{\tf}{\mathscr{f}}
\newcommand{\tm}{\mathscr{m}}

\newcommand{\stb}{\mathsf{b}}

\newcommand{\tpiI}{(2 \pi \ri)}
\newcommand{\ma}{m_\alpha}
\newcommand{\LC}{D}

\newcommand{\QB}{\boldsymbol{Q_B}}
\newcommand{\tQB}{\boldsymbol{\tilde{Q}_B}}
\newcommand{\bk}{\boldsymbol{k}}

\newcommand{\Zk}{Z_{\bk}}
\newcommand{\Zmodk}{{\cal Z}_{\bk}}
\newcommand{\Fk}{F_{\bk}}
\newcommand{\Fmodk}{{\cal F}_{\bk}}

\newcommand{\bt}{\boldsymbol{t}}
\newcommand{\bm}{\boldsymbol{m}}
\newcommand{\bd}{\boldsymbol{d}}
\newcommand{\bQ}{\boldsymbol{Q}}
\newcommand{\Qm}{\boldsymbol{Q_m}}
\newcommand{\dm}{\boldsymbol{d}_m}
\newcommand{\Res}{\mbox{Res}}

\def\inttorus{\oint\displaylimits_{\begin{tikzpicture}
		\node [draw, trapezium, minimum width=0.5cm, trapezium left angle=60, trapezium right angle=120] {} ;
		\end{tikzpicture}}}

\usepackage{colortbl}
\usepackage{booktabs}                       

\def\({\left(}
\def\){\right)}
\def\[{\left[}
\def\]{\right]}

\def\md{\mathbf}

\def\e{\md e}

\def\Mkn{M_{\bk,n}}
\def\tMki{\mathscr{m}_{\bk,i}}

\def\mki{m_{\bk,i}}

\def\t{t}

\def\indexm{i_{\bm}}
\def\indexz{i_z}

\def\indexL{i_L}
\def\indexR{i_R}
\def\indexti{i_{t_\iota}}
\def\index1{i_1}    

\newcommand{\ri}{{\rm i}}

\begin{document}
  \begin{titlepage}
    {}~ \hfill\vbox{ \hbox{} }\break
	
    \rightline{LPTENS/18/13}

    \vskip 1 cm
	
    \vskip 1 cm
    
    \begin{center}
    	\Large \bf Computing the  elliptic genus of higher rank E-strings from genus 0 GW invariants
    \end{center}
    
    \vskip 0.8 cm

    \centerline{Zhihao Duan, Jie Gu,  Amir-Kian Kashani-Poor} 
	
    \vskip 0.2in
    \begin{center}{\footnotesize
      \begin{tabular}{c}
      {\em LPTENS, CNRS, PSL University, Sorbonne Universit\'{e}s, UPMC, 75005 Paris, France}\\[0ex]
      \end{tabular}
    }\end{center}
		
    \setcounter{footnote}{0}
    \renewcommand{\thefootnote}{\arabic{footnote}}
    \vskip 60pt
    \begin{abstract} 
      {}{\bf Abstract:} We show that the elliptic genus of the higher rank E-strings can be computed based solely on the genus 0 Gromov-Witten invariants of the corresponding elliptic geometry. To set up our computation, we study the structure of the topological string free energy on elliptically fibered Calabi-Yau manifolds both in the unrefined and the refined case, determining the maximal amount of the modular structure of the partition function that can be salvaged. In the case of fibrations exhibiting only isolated fibral curves, we show that the principal parts of the topological string partition function at given base-wrapping can be computed from the knowledge of the genus 0 Gromov-Witten invariants at this base-wrapping, and the partition function at lower base-wrappings. For the class of geometries leading to the higher rank E-strings, this leads to the result stated in the opening sentence.

    \end{abstract}
		
    {\let\thefootnote\relax
      \footnotetext{duan@lpt.ens.fr, jie.gu@lpt.ens.fr, kashani@lpt.ens.fr}}
	
  \end{titlepage}
  \vfill \eject

  \tableofcontents
  \newpage
    
\section{Introduction} 

A recent revival of interest in six dimensional quantum field theories has led to a better understanding of the topological
string partition function $Z_{\text{top}}$ on an elliptically fibered Calabi-Yau threefold $X$ over a complex base $B$ \cite{Huang:2013,Haghighat:2014,Kim:2014,Haghighat:2014vxa,HKK,Kim:2015jba,Gadde:2015tra,Shabbir:2015oxa,Kim:2016foj,DelZotto:2016pvm,Hohenegger:2016yuv,Gu:2017ccq,Hayashi:2017jze,Haghighat:2017vch,Bastian:2017ing,DelZotto:2017mee,Kim:2018gak,Lee:2018urn}. This has resulted in modular expressions for the coefficients $\Zmodk(\tau,z,\bm)$ of the expansion of
$Z_{\text{top}}$ in suitably shifted exponentiated K\"ahler moduli $\tQB$ of base classes,
\begin{equation} \label{Zk}
  Z_{\text{top}} = Z_0 \cdot \(1 + \sum_{\bk \neq 0} \Zmodk\tQB^{\bk}\)
  \,.
\end{equation}
Here and throughout this note, we use the notation $\bk = (k_1,k_2,\ldots)$, $k_i \ge 0$, to denote a curve class in the1
base $B$. $\Zmodk(\tau,z,\bm)$ is 
a Jacobi form whose modular parameter is the K\"ahler modulus $\tau$
of the elliptic fiber, and with elliptic parameters the string
coupling $z = \tfrac{g_s}{2\pi}$, as well as the K\"ahler
moduli $\bm$ of the fibral curve classes.

The Gopakumar-Vafa form \cite{GV1,GV2} of $Z_{\text{top}}$ reveals that
$\Zmodk$ must exhibit poles; $\Zmodk$ as a Jacobi form must hence be
meromorphic. Unlike the ring of weak Jacobi forms, whose elements are holomorphic, the ring of meromorphic Jacobi forms is not finitely generated. In
\cite{HKK,Gu:2017ccq,DelZotto:2017mee,Kim:2018gak}, progress hinged on
expressing $\Zmodk$ as a quotient of weak Jacobi forms, 
\begin{equation} \label{eq:general_form}
\Zmodk = \frac{\sum c_i \phi_{\bk,i}(\tau,z,\bm)}{\phi^D_{\bk}(\tau,z,\bm)} \,.
\end{equation}
The denominator takes a universal form depending only on the knowledge of the classical intersection numbers of the divisors of the elliptically fibered Calabi-Yau manifold $X$. This data also fixes the weight and indices of the numerator, allowing an expansion in appropriate ring generators. The expansion coefficients $c_i$ must be determined by imposing additional constraints on $\Zmodk$. 

In this note, we show that in the case of the higher rank $E$-strings, genus 0 Gromov-Witten invariants provide sufficient boundary conditions to determine the $c_i$, in other words, to compute the topological string to all orders in $g_s$.

Our analysis relies on determining the principal parts (the negative degree terms in the Laurent expansion) of $\Zmodk$ around all of its poles. We achieve this by combining modular and multi-wrapping properties of $\Ztop$. While the modular properties are manifest in the form \eqref{eq:general_form}, the multi-wrapping properties, as captured by the Gopakumar-Vafa formula, are formulated more naturally in terms of the topological string free energy $\Fmodk$, related to $\Zmodk$ via 
\begin{equation}\label{Fk}
Z_{\text{top}} = Z_0 \cdot \exp \(\sum_{\bk \neq 0} \Fmodk \tQB^{\bk}\) \,.
\end{equation}
We thus begin our analysis by asking to what extent the modularity properties of $\Zmodk$ carry over to $\Fmodk$. In a nutshell, the answer is that the string coupling as elliptic parameter is lost. However, the coefficients of $\Fmodk$ in a $g_s$ expansion are elements of the ring of Jacobi forms with modular parameter $\tau$ and elliptic parameters $\bm$, tensored with the ring of quasi-modular forms. We next study the pole structure of $\Fmodk$ as a function of $z$. In the case of elliptic fibrations with only isolated fibral curves (geometries that engineer SCFTs in F-theory with at most abelian gauge symmetry are in this class), the denominator $\phi^D_{\bk}$ in \eqref{eq:general_form} in fact does not depend on fibral K\"ahler classes. This gives rise to a tractable pole structure for the free energy, and ultimately allows us to reduce the computation of all negative index Laurent coefficients of $\Zmodk$ to the knowledge of genus 0 Gromov-Witten invariants. The analogous considerations in the case of more general elliptic fibrations fail, as we explain.

Our analysis also applies to the case of the refined topological string. Here, the elliptic parameter $z = \tfrac{g_s}{2\pi}$ is replaced by two elliptic parameters $z_{1,2} = \tfrac{\epsilon_{1,2}}{2\pi}$. Even though the refined coupling $s = \epsilon_1 + \epsilon_2$ behaves in many ways as a K\"ahler parameter of a non-compact divisor (hence as a mass parameter in the geometrically engineered field theory), see e.g. \cite{Huang:2011qx}, it cannot be salvaged as an elliptic parameter at the level of the free energy, as we explain; expansion in both $z_1$ and $z_2$ is required. The role of the genus 0 Gromov-Witten invariants in determining the higher rank E-string partition function is played in the refined case by the Nekrasov-Shatashvili data
\begin{equation} 
n_g^{\bd,\text{NS}} = \sum_{g_L+g_R = g}n_{g_L,g_R}^{\bd} \ ,
\end{equation}
with $n_{g_L,g_R}$ indicating the refined Gopakumar-Vafa invariants.

The structure of this paper is the following. In section \ref{s:cycleclasses}, we review the classes of 4-cycles that occur in elliptically fibered Calabi-Yau manifolds, and explain the role the corresponding K\"ahler classes play in $\Zmodk$. We study the modularity and pole structure of the topological free energy $\Fmodk$ in section \ref{s:structureF}. Section \ref{s:computing_laurent_Z} explains how to use this information to reduce the computation of the principal parts of $\Zmodk$ in the case of elliptic fibrations with only isolated fibral curves to the knowledge of genus 0 Gromov-Witten invariants. Finally, we explain in section \ref{s:nidx} in which cases the principal parts of $\Zmodk$ determine it completely, and explain the relation to determining the expansion coefficients $c_i$ in \eqref{eq:general_form} by imposing vanishing conditions on Gopakumar-Vafa invariants.

\section{Classes of 4-cycles on elliptic fibrations and how they enter in
  $Z_{\text{top}}$} \label{s:cycleclasses}

	\subsection{Classes of 4-cycles}
        
	Compact elliptically fibered Calabi-Yau manifolds can be constructed as hypersurfaces in a projective bundle over a compact K\"ahler base manifold $B$. The hypersurface is cut out via a Weierstrass equation in variables $(x:y:z)$ which are sections of the projective bundle; the coefficients $f$ and $g$ are sections of appropriate negative multiples of the canonical bundle of $B$, chosen to impose the Calabi-Yau condition on the hypersurface. The singularity structure of the elliptic fiber depends on the vanishing properties of these sections, and is hence intimately tied to the divisor structure of the base $B$. Non-compact models can be obtained via degeneration limits of this setup. For instance, decompactification along the fiber direction can lead to local Calabi-Yau manifolds which are the total space of the canonical bundle of a surface, while decompactification perpendicular to the fiber direction yields elliptic fibration over a non-compact surface, such as the geometries appearing as building blocks in the classification scheme of 6d SCFTs via F-theory \cite{Heckman:2015bfa}.

	The topological string partition function depends on the topological string coupling constant $g_s$ (or, in the case of refinement, on two parameters $\epsilon_{1,2}$ which can be organized as $g_s^2 = - \ep1 \eps2$ and $s = (\ep1 + \eps2)^2$) and (in the A-model perspective) on K\"ahler parameters associated to homology classes in $H_2(X)$ of the Calabi-Yau manifold $X$. In the generic Gopakumar-Vafa formula, which we will review in the next section, all K\"ahler parameters enter the partition function on the same footing. When $X$ is elliptically fibered, different K\"ahler parameters are distinguished by the action of the monodromy group on the associated curve classes  \cite{HKK}. The curve classes in the base $B$ are essentially invariant under this action (see the discussion around equations \eqref{eq:shifted_K_parameters_1} and \eqref{eq:shifted_K_parameters_2} for the precise statement). The associated K\"ahler parameters are treated as expansion parameters. The coefficients $\Zmodk$ of the expansion in appropriately shifted exponentiated base classes, with $\bk$ indicating the base class,  are Jacobi forms. All remaining K\"ahler parameters as well as the string coupling play the role of the modular and elliptic parameters of these Jacobi forms.

	To understand the roles played by different curve classes in the topological string partition function on an elliptic Calabi-Yau threefold, we focus on the case of compact Calabi-Yau manifolds, and argue, via Poincar\'e duality, in terms of 4-cycles rather than 2-cycles. The non-compact case can then be obtained via degeneration. We can distinguish between 4 classes of divisors \cite{MR2041769,Morrison:2012ei}:
	\begin{enumerate}
		\item The pullbacks $B_\alpha$ of divisors (curves) of the base $B$ to $X$ via the projection $\pi: X \rightarrow B$, i.e.\ $B_\alpha = \pi^* H_\alpha$.
		\item The zero section of the fibration, which is topologically the base $B$.
		\item Divisors $T_{\kappa,I}$ consisting of fiber components $\alpha_{\kappa,I}$ arising from the resolution of singularities of the fibration over curves $b_\kappa$ in $B$, fibered over $b_\kappa$.
		\item Divisors $S_i$ associated to other than the zero section of the fibration.
	\end{enumerate}
	By \cite{MR2041769}, all 4-cycles in $X$ fall into one of these four classes. Upon replacing the 4-cycles $S_i$ by their images under the threefold version of the Shioda map \cite{Park:2011ji}, the Poincar\'e dual 2-cycles to these 4 classes are
	\begin{enumerate}
		\item Curve classes $H_\alpha$ of the base $B$.
		\item $f$, the fiber class of the fibration.
		\item Fiber components $\alpha_{\kappa,I}$.
		\item Isolated rational curves $s_i$ in the fiber.
	\end{enumerate}
 In the F-theory compactification on
        $X $, curve classes of type 1 give rise to tensor multiplets,
        those of type 3 to non-abelian vector multiplets, while those
        of type 4 give rise to abelian vector multiplets as well as
        hypermultiplets charged under them.

	The corresponding K\"ahler classes play different roles in $Z_{top}$, depending on the monodromy action on the associated curve classes: 
	\begin{enumerate}
		\item The exponentials $\QB=(Q_{1}, \ldots, Q_{b_2(B)})$ of the base K\"ahler classes $t_\alpha = \int_{H_\alpha} J$ can be rendered invariant upon an appropriate shift explained below (see equations \eqref{eq:shifted_K_parameters_1} and \eqref{eq:shifted_K_parameters_2}). $Z_{\text{top}}$ is expanded in terms of the shifted variables $\tQB$, see \eqref{Zk}, to yield the Jacobi forms $\Zmodk$ as coefficients.
		\item   $\tau = \int_f J$ is the modular parameter of the Jacobi forms $\Zmodk$.
		\item $c_{\kappa,I} = \int_{\alpha_{\kappa,I}} J$ are elliptic parameters on which the position of the poles of $\Zmodk$ as a function of $g_s$ (or $\epsilon_{1,2}$ in the refined case) depends.\footnote{Note that $F$-theory, and therefore the 6 dimensional theory obtained by compactifying on such geometries, is not sensitive to these K\"ahler parameters. Upon compactifying on a circle, however, the resulting vector multiplets exhibit real scalar fields whose VEV keeps track of the blow-up cycle size, and which collectively transform under the global symmetry given by the Weyl group of the gauge group.}
		\item $m_i = \int_{\sigma(S_i)} J$ give rise to elliptic parameters which do not modify the position of the poles of $\Zmodk$ as a function of $g_s$ (or $\epsilon_{1,2}$ in the refined case).
	\end{enumerate} 
	
	We refer to the curve classes of type 3 and 4 as fibral, and denote the corresponding K\"ahler parameters collectively as $t_{\iota}$.

	\subsection{The base degree $\bk$ partition function as Jacobi form}
	We now turn to the discussion of the structure of the partition function at base-wrapping degree $\bk$, $\Zmodk$. At the end of the section, we point out the modifications necessary to generalize to the case of refinement.
	
    One path to the identification of the transformation properties of $\Zmodk$ runs via the holomorphic anomaly equations of the topological string \cite{BCOV,Hosono:1999qc,HKK}. These can be rewritten in the form
       \be \label{eq:holanom}
       \left(\frac{\partial}{\partial E_2} + \frac{\pi^2}{3} M_{\bk} \right) \Zmodk = 0 \,.
       \ee
    $M_{\bk}$, dubbed the index bilinear form in \cite{DelZotto:2017mee}, depends quadratically on the string coupling $z = \tfrac{g_s}{2\pi}$ and all fibral K\"ahler parameters,
    \be
    M_{\bk} = \indexz(\bk) z^2 + \sum \indexti(\bk) t_\iota^2 .
    \ee
    It is easy to show that a convergent power series
    \be
    \sum_{\ell_z,\boldsymbol{\ell}} a_{\ell_z,\boldsymbol{\ell}} z^{2 \ell_z} \bt^{2\boldsymbol{\ell}}
    \ee
    with coefficients $a_{\ell_z,\boldsymbol{\ell}}$ of weight $w +2(\ell_z + \sum_{\iota} \ell_{\iota})$ in the ring $\IC[E_2,E_4,E_6]$ of quasi-modular forms  which satisfies the differential equation \eqref{eq:holanom} is a weak Jacobi form with elliptic parameters $z$, $t_{\iota}$ of index $\indexz(\bk)$, $\indexti(\bk)$ respectively.
    
    The program of rewriting the holomorphic anomaly equations of \cite{BCOV} in the form \eqref{eq:holanom} has been carried out for local 1/2 K3 in \cite{Hosono:1999qc} and for the elliptic fibration over the base $B=\IF_1$ in \cite{Klemm:2012sx}, though some details remain to be ironed out.
    
    A currently more efficient path towards determining the indices
    $\indexti$ for all fibral curve classes proceeds via F-theory
    compactifications on the elliptically fibered manifold $X$. The
    resulting 6d theory exhibits non-critical strings which arise via
    D3 branes wrapping curves $C_{\bk}$ in the base manifold
    $B$. $\Zk$ for $\bk \neq 0$ can be identified with the elliptic
    genus of these strings \cite{Klemm:1996,Haghighat:2014vxa}. The
    transformation properties of elliptic genera under modular
    transformations have been argued for in
    \cite{Benini:2012cz,Benini:2013cda}. One can use this vantage
    point to fix the index bilinear $M_{\bk}$ in terms of the anomaly
    polynomial of the worldsheet theory\footnote{See \cite{DelZotto:2018tcj} for an analysis of these worldsheet theories beyond the minimal SCFTs.} of these strings
    \cite{DelZotto:2016pvm,Gu:2017ccq,DelZotto:2017mee,Lee:2018urn}, which has been computed in \cite{Kim:2016foj, Shimizu:2016lbw}. For our purposes, the
    important characteristic is that all indices other that $\indexz$
    depend linearly on the base-wrapping degree $\bk$; $\indexz$,
    which takes the elegant form \be \indexz(\bk) = \frac{C_{\bk}
      \cdot (C_{\bk} + K_B)}{2} \,, \ee is quadratic in $\bk$.

	An important ingredient in solving for $\Zmodk$, once its transformation properties under the modular group have been identified, is determining its pole structure. The Gopakumar-Vafa form of the free energy motivates the ansatz
	\be \label{eq:denom_no_gauge}
	\phi_{\bk}^D = \prod_{i=1}^{b_2(B)}\prod_{s=1}^{k_i} \phi_{-2,1}(\tau,sz)
	\ee
	 for the denominator in \eqref{eq:general_form}. In the absence of curves of type 3, this ansatz has been verified for numerous examples in \cite{HKK,Gu:2017ccq}. Indeed, isolated rational curves, such as the curves of type 4, are locally modeled by the conifold. Lacking a moduli space, they are not expected to give rise to a contribution (hence $\bm$ dependence) in the denominator of $\Ztop$. On the other hand, rational curves of self-intersection number less than $-1$, such as the curves of type 3, do necessarily exhibit a moduli space, and are expected to modify the pole structure of $\Ztop$ (cf. the discussion in section 3.4 of \cite{Iqbal:2004ne} juxtaposing $-1$ and $-2$ curves). This expectation is born out by localization computations \cite{DelZotto:2016pvm,Lee:2018urn} and topological string computations \cite{Gu:2017ccq, DelZotto:2017mee,Lee:2018urn}, the latter primarily in the case of local geometries.
	
	To describe how the ansatz \eqref{eq:denom_no_gauge} must be modified in the presence of curves of type 3, recall that these arise upon resolution of singularities over a divisor $C$ in the base $B$. The intersection matrix of the fiber components above $C$ is captured by the negative Cartan matrix of an associated affine Lie algebra $\hat{\mathfrak{g}}$.  Specifically, $X$ contains an elliptic surface with base $C$ and reducible fiber
	\be
	F = F_0 + \sum_i a_i^\vee E_i \,.
	\ee
	The $E_i$ here are the exceptional curves that resolve the singularity. $a_i^\vee$ are the comarks of $\mathfrak{g}$. There is a monodromy action on the curves $\{E_i\}$ captured by the Weyl group of $\mathfrak{g}$. To make this manifest, the K\"ahler parameters $m_i$ associated to a choice of basis of the exceptional curves are assembled into an element of the complexified root lattice $\bm = \sum_i m_i \boldsymbol{\omega_i}$, the $\omega_i$ denoting the fundamental weights of $\mathfrak{g}$.

	The denominator in \eqref{eq:general_form} in the case $b_2(B)
        =1$ given in \cite{DelZotto:2017mee} reduces in the unrefined case (up to a sign) to 
	\begin{equation} \label{eq:denom-gauge}
	\prod_{s=1}^k\(
	\phi_{-2,1}(\tau,sz)
        \prod_{\ell=0}^{s-1}\prod_{\alpha\in\Delta_+}
        \phi_{-2,1}\(\tau,(s-1-2\ell)z+m_{\alpha}\)\)
	\ ,
	\end{equation}
	where we have defined
	\begin{equation}
	m_\alpha = (\bm, \alpha^\vee) 
	\end{equation}
	for any root $\alpha \in \Delta$. The naive generalization beyond $b_2(B)=1$ should be correct, but no computations have yet been performed in this case.
	
	Finally, we explain the relation between the exponentiated expansion parameters $\tQB$ and the exponentiated base K\"ahler classes $\QB$. As argued in \cite{HKK}, based on genus 0 observations in \cite{Candelas:1994}, for the case of non-singular elliptic fibrations, the base class shifted by an appropriate multiple of the fiber class is invariant under the $SL(2,\IZ)$ monodromy action, up to a sign corresponding to the same multiplier system as an appropriate power of $\eta^{12}(\tau)$. The modular properties of $\Ztop$ are thus manifest when expanding in
	\be \label{eq:shifted_K_parameters_1}
	\tQB = \left(\frac{\sqrt{q}}{\eta^{12}(\tau)}\right)^{-C_{\bk}\cdot K_B}   \QB \,.
	\ee
	From the identification of the topological string with the elliptic genus of the worldsheet theory of non-critical strings, this result was generalized to the case of singular fibrations in \cite{Haghighat:2014vxa} to
	\be \label{eq:shifted_K_parameters_2}
	\tQB = \left(\frac{\sqrt{q}}{\eta^{12}(\tau)\prod_{i=1}^r Q_i^{a^{\vee}_i} }\right)^{-C_{\bk}\cdot K_B}   \QB \,.
	\ee
	Note that the appropriate expansion parameter to extract the enumerative data encoded in $\Ztop$ remains $\QB$; this is also the parameter that occurs in the Gopakumar-Vafa presentation of the free energy which we shall review in section \ref{s:structureF}. We thus introduce straight letters $Z$ and $F$ to denote the corresponding expansion coefficients, such that
	\be \label{eq:enum_to_mod}
	\Zk = \left(\frac{\sqrt{q}}{\eta^{12}(\tau)\prod_{i=1}^r Q_i^{a^{\vee}_i}}\right)^{-C_{\bk}\cdot K_B}  \Zmodk \,, \quad	\Fk = \left(\frac{\sqrt{q}}{\eta^{12}(\tau)\prod_{i=1}^r Q_i^{a^{\vee}_i}}\right)^{-C_{\bk}\cdot K_B}  \Fmodk \,,
	\ee
	where the product in the denominator ranges over divisors of the geometry of type 3.
	
The refined version of the results discussed in the section is also known \cite{DelZotto:2016pvm,Gu:2017ccq,DelZotto:2017mee}. Introducing the variables
\be
z_{1,2} = \frac{\epsilon_{1,2}}{2\pi} \quad \,, \quad \quad z_{L,R} = \frac{\epsilon_{L,R}}{2\pi}  \,,
\ee
the refined partition function takes the general form 
\begin{equation} \label{eq:general_form_ref}
\Zmodk = \frac{\sum c_i \phi_{\bk,i}(\tau,z_L,z_R,\bm)}{\phi^D_{\bk}(\tau,z_1,z_2,\bm)} \,,
\end{equation}
with $z_{L,R}$ (or equivalently $z_{1,2}$) serving as elliptic parameters. The denominator in the absence of divisors of type 3 is given by \cite{DelZotto:2016pvm,Gu:2017ccq}
\be \label{eq:denom_no_gauge_ref}
\phi_{\bk}^D = \prod_{i=1}^{b_2(B)}\prod_{s=1}^{k_i} \phi_{-1,1/2}(\tau,s z_1)\phi_{-1,1/2}(\tau,s z_2) \,,
\ee
where $\phi_{-1,1/2}^2 = \phi_{-2,1}$. When such divisors are present, the denominator (in the case $b_2(B)
=1$) is \cite{DelZotto:2016pvm,Kim:2018gak}
\be
\prod_{s=1}^k \prod_{j=\pm 1} \left(\phi_{-1,1/2}(s (z_R + j z_L) )  \prod_{\ell=0}^{s-1} \prod_{\alpha \in \Delta_+} \phi_{-1,1/2} ((s+1) z_R + (s-1-2\ell) z_L + j m_\alpha ) \right) \,.
\ee
Note that this specializes to \eqref{eq:denom-gauge} for $z_R = 0, z=z_L$ up to an irrelevant sign.

\section{The structure of the topological string free energy} \label{s:structureF}
\subsection{Setting up the analysis}
Depending on the computational approach one takes, the topological string partition function or its free energy moves to the fore. The
considerations in section \ref{s:cycleclasses} centered around $Z_{\text{top}}$, as this quantity has the better transformation
properties under the modular group. From the Gopakumar-Vafa presentation, it is clear however
that the partition function contains redundant information due to multi-wrapping contributions. This redundancy is easily identified at
the level of the free energy. It takes the general form 
\begin{equation}  \label{fgvtop}
F(g_s,\bt)=\sum_{w\geq 1\atop g \geq 0} \sum_{\bd\in H_2(X,\mathbb{Z})} n^{\bd }_g
\left(2 \sin \left( \frac{w g_s }{2}\right)\right)^{2 g -2} \frac{\bQ^{w \bd}}{w}   \,,\quad \quad n^{\bd}_g\in \mathbb{Z}\,,
\end{equation}
with single-wrapping contributions
 \begin{equation}\label{eq:fn}
f(g_s,\bt) = \sum_{g \geq 0} \sum_{\bd\in H_2(X,\mathbb{Z})} n^{\bd}_g
\left(2 \sin \left( \frac{g_s}{2}\right)\right)^{2 g -2} \bQ^{\bd}   \,.
\end{equation}

We thus turn to the study of the free energy on elliptic
Calabi-Yau manifolds in this section. These considerations will play
an important role when determining the interdependence of the
principal parts of the base degree $\bk$ contributions $Z_{\bk}$ to
the partition function in section \ref{s:computing_laurent_Z}.
    
As in the case of the partition function, it will be convenient to consider coefficients $F_{\bk}$ of the
free energy in an expansion in base degree classes $\QB$ (recall that $z = \tfrac{g_s}{2\pi}$),
\begin{equation}
  Z_{\text{top}} = Z_0 \left(\sum_{\bk>0} Z_{\bk} \QB^{\bk}\right) = Z_0
  \exp \left( \sum_{\bk>0} F_{\bk} \QB^{\bk} \right) \,,
\end{equation}
with single-wrapping contribution

  \be
 f_{\bk}(z,\bt) = \sum_{g,d_{\tau},\dm} n_g^{\bk,d_{\tau},\dm }(2\sin \pi z)^{2g-2} q^{d_\tau} \Qm^{\dm} \,.
\end{equation}
We can invert
\begin{equation} \label{eq:ZkfromF}
  Z_{\bk} = \sum_{n=1}^\infty
  \frac{1}{n!} \sum_{\bk_1, \ldots, \bk_n > \boldsymbol{0} \atop \sum \bk_i = \bk}
  \prod_{i=1}^n F_{\bk_i}
\end{equation}
to obtain the free energy at base wrapping $\bk$ in terms of
$Z_{\bk'}$, $|\bk'| \le |\bk|$,
\begin{equation} \label{eq:FfromZ}
  F_{\bk} = \sum_{n=1}^\infty
  \frac{1}{n!}
  \sum_{\bk_1, \ldots, \bk_n > \boldsymbol{0} \atop\sum \bk_i = \bk}
  a^{(\bk)}_{\bk_1,\ldots,\bk_n} \prod_{i=1}^n
  Z_{\bk_i} = \sum_n \Mkn(\{Z_{\bk'}:|\bk'| \le |\bk|\})\,.
\end{equation}
Note that the sums over $n$ in \eqref{eq:ZkfromF} and
\eqref{eq:FfromZ} are effectively finite due to the constraint
$\bk_i > \boldsymbol{0}$ on the $n$ summation coefficients
$\bk_1, \ldots, \bk_n$. This constraint also implies that the
monomials $\Mkn$ are homogeneous in the index $\bk_i$ of is arguments,
$\sum_i \bk_i = \bk$.

The integer coefficients $a^{(\bk)}_{\bk_1,\ldots,\bk_n}$ are easily computed iteratively. In particular, $a^{(\bk)}_{\bk} = 1$. We give the corresponding monomial the index 1,
\be
M_{\bk,1} = Z_{\bk} \,.
\ee
In the case of $b_2(B)= 1$, e.g., we have
\begin{equation}\label{FfromZr1}
  F_1 =  Z_1 \ ,\quad
  F_2 =  Z_2 - \frac{1}{2}Z_1^2 \ ,\quad
  F_3 =  Z_3 - Z_1 Z_2 + \frac{1}{3}Z_1^3  \ ,\quad \ldots
\end{equation}
and 
\be
M_{1,1} =  Z_1 \ ,\quad
\ee
\be
M_{2,1} = Z_2 \,, \quad M_{2,2}=- \frac{1}{2}Z_1^2 \ ,
\ee
\be
M_{3,1} = Z_3\,, \quad M_{3,2} =  - Z_1 Z_2 \,,\quad M_{3,3} = \frac{1}{3}Z_1^3  \ ,\quad\ldots \,.
\ee

To compute the single-wrapping contribution $f_{\bk}$, it suffices to subtract from $F_{\bk}$ appropriate linear combinations of $F_{\bk'}$, $|\bk'| < |\bk|$, evaluated at integer multiples of all of their arguments. Symbolically, we write this as
\begin{equation}
f_{\bk}(z,\bt) = F_{\bk}(z,\bt) + \pFf(F_{|\bk'|<|\bk|}(*z,*\bt)) \ ,
\end{equation}
where $*$ is a placeholder for a possible multi-wrapping factor, and $\pFf$ is a polynomial with the property that every contributing
monomial $\prod_i F_{\bk_i}(w_i z,w_i\bt)$ satisfies
$\sum_i w_i \bk_i = \bk$. In the case $b_2(B)=1$, the first few
expressions are
\begin{equation}
\begin{aligned}
f_1(z,\bt) &= F_1(z,\bt)\,,\\
f_2(z,\bt) &= F_2(z,\bt) -
\frac{1}{2}F_1(2z,2\bt)  \,,\\
f_3(z,\bt) &= F_3(z,\bt) -
\frac{1}{3}F_1(3 z,3\bt) \\
 & \ldots  \,.
\end{aligned}
\end{equation}

Combining this with \eqref{eq:FfromZ}, we immediately obtain
\begin{equation}\label{eq:fZ-ref}
  f_{\bk}(z,\bt) = Z_{\bk}(z,\bt) +
  \pZf(Z_{|\bk'|<|\bk|}(*z,*\bt)) \ .
\end{equation}
Again in the case $b_2(B)=1$, the first few relations are
\begin{equation} \label{eq:f_to_Z_r1}
\begin{aligned}
f_1(z,\bt) &= Z_1(z,\bt) \ ,\\
f_2(z,\bt) &= Z_2(z,\bt) - \frac{1}{2}Z_1(2z,2\bt)
- \frac{1}{2} Z_1(z,\bt)^2 \ , \\
 f_3(z,\bt) &= Z_3(z,\bt) - \frac{1}{3}Z_1(3z,3bt)
 - \frac{1}{6} Z_1(z,\bt)^3 \ , \\
 & \ldots .
\end{aligned}
\end{equation}
We of course can also invert these relation to obtain
\begin{equation}\label{eq:Zf-ref}
  Z_{\bk}(z,\bt) = f_{\bk}(z,\bt) +
  \pfZ(f_{|\bk'|<|\bk|}(*z,*\bt)) \,.
\end{equation}
When $b_2(B)=1$, 
\begin{equation} \label{eq:Z_to_f_r1}
\begin{aligned} 
Z_1(z,\bt) &= f_1(z,\bt) \\
Z_2(z,\bt) &= f_2(z,\bt) +
\frac{1}{2}f_1(2 z,2\bt) +
\frac{1}{2}f_1(g_s,\bt)^2 \\
Z_3(z,\bt) &= f_3(z,\bt) +
\frac{1}{3}f_1(3 z,3\bt) +
\frac{1}{6}f_1(z,\bt)^3 \\
 &\ldots \,.
\end{aligned}
\end{equation}

In section \ref{s:computing_laurent_Z}, it will be more natural to reorder \eqref{eq:fZ-ref},
\begin{equation}\label{eq:rec-ref}
  Z_{\bk}(z,\bt) = f_{\bk}(z,\bt) -
  \pZf(Z_{|\bk'|<|\bk|}(*z,*\bt)) \,.
\end{equation}
This equation encodes the fact that when increasing the base degree by one step to $\bk$, all the new information required to reconstruct $Z_{\bk}$ is captured by the single-wrapping contribution to the free energy $f_{\bk}$. Anticipating our discussion in section \ref{s:computing_laurent_Z}, we also introduce the monomials $\mki$ constituting $\pZf$,
\be \label{eq:rec-ref_aux}
	Z_{\bk} = f_{\bk} + \sum \mki \,,
\ee
where each $\mki$ is of the form
\be \label{eq:constraint_sj}
\mki \propto \prod_j Z_{\bk_{i,j}} (s_j \tau, s_j z , s_j \bm) \,, \quad \sum_j s_j \bk_{i,j} = \bk \,.
\ee

All of these formulae generalize straightforwardly to the refined case, given the refined Gopakumar-Vafa expansion \cite{Hollowood:2003cv}
\begin{equation}\label{GVref}
F(\epsilon_{L,R},\bt) = \sum_{g_{L,R} \geq 0}\sum_{w\geq 1}\sum_{\bd\in H_2(X,\IZ)}
\frac{n^{\bd}_{g_L,g_R}}{w}\frac{(2\sin\tfrac{w\epsilon_L}{2})^{2g_L}(2\sin\tfrac{w\epsilon_R}{2})^{2g_R}}{2\sin
	\tfrac{w(\epsilon_R+\epsilon_L)}{2}2\sin
	\tfrac{w(\epsilon_R-\epsilon_L)}{2}} \bQ^{w \bd} \ ,\quad n^{\bd}_{g_L,g_R}\in \IZ 
\end{equation}
as starting point. In terms of the equivariant parameters $\epsilon_{1,2}$, we have $\epsilon_{L,R} = \tfrac{\epsilon_1 \mp \epsilon_2}{2}$. The conventional (non-refined) topological string is obtained by setting $\epsilon_2 = -\epsilon_1$, $g_s^2 = \epsilon_1^2$.

	\subsection{Transformation behavior}
	While the object with the best transformation behavior under the modular group is $\Zmodk$, the partition function at fixed base-wrapping, some of this behavior survives to the level of the free energies $\Fmodk$. Perhaps somewhat surprisingly, the unrefined and the refined situations are qualitatively different. We discuss these two cases separately in this subsection.
	
	\subsubsection{The unrefined case} \label{ss:transf_unref}

	From the relation  \eqref{eq:FfromZ} between free energy and
        partition function, we can conclude that $\Fmodk$ has the
        general form 
	\be \label{eq:free_energy_gen_form}
	\Fmodk(\tau,z,\bm) = \sum_n \Mkn({\cal Z}_{\bk'}:|\bk'| \le |\bk|\}) \,.
	\ee
	Note that the $\Mkn({\cal Z}_{\bk'}:|\bk'| \le |\bk|\})$ as
        monomials in Jacobi forms are again Jacobi forms. They have
        identical $\bm$-index, as this quantity is  linear in the
        base-wrapping degree $\bk$, and the $\Mkn$ are homogeneous
        with regard to base-wrapping. By contrast, the $z$-index
        depends quadratically on the base-wrapping degree, hence
        varies with the index $n$. $\Fmodk$ is hence not a Jacobi form of fixed $z$-index. To make progress, we Laurent expand in the variable $z$,
	\be
	\Fmodk(\tau,z,\bm) = \sum_{g=0}^\infty \cF_{\bk,g}(\tau,\bm) \,z^{2g-2} \,.
	\ee
	The coefficients $\cF_{\bk,g}(\tau,\bm)$ are the genus $g$ contributions to the free energy at base degree $\bk$. 
	
	In the absence of divisors of type 3 (i.e. non-abelian gauge symmetry in the case of SCFTs), $\Zmodk$ is of the form 
	\begin{equation}
	\Zmodk = \frac{\sum c_i \phi_{\bk,i}(\tau,z,\bm)}{\phi^D_{\bk}(\tau,z)} \,,
	\end{equation}
	with $\phi_{\bk,i}(\tau,z,\bm)$ an element of the tensor product of the ring of Jacobi forms in $z$ and in $\bm$, and with $\phi^D_{\bk}(\tau,z)$ having $\tau$-independent leading coefficient in $z$. The expansion in $z$ thus yields Jacobi forms in $\bm$ with the Taylor coefficients of Jacobi forms in $z$, i.e. quasi-modular forms (see appendix \ref{s:app_jac}), as coefficients. By the argument above, the $\bm$-index of $\cF_{\bk,g}$ is equal to that of $\Zmodk$. Restoring the $\eta$-dependence, the weight of $F_{\bk,g}$ can be read off from the power of $z$ it multiplies; it is equal to $2g-2$.
	
	The structure of $\cF_{\bk,g}$ for these cases can be further
	constrained. Let us restrict for simplicity to the case $b_2(B)=1$. Fix $k$ and $g$. The contributions of the second
	Eisenstein series $E_2(\tau)$, the source of the ``quasi'' in the
	quasi-modularity of $\cF_{\bk,g}$, stem from the Taylor expansion of the
	Jacobi forms $A$ and $B$ in $z$. From the Taylor series of $A$ and
	$B$, given in \eqref{ABquasimodular}, we can infer that the highest
	power of $E_2$ at given order in $z$ will arise from the contribution
	to $\cF_{\bk,g}$ with the highest power in $B$. At genus $g$, this
	highest power is $2k+2g-2$; the contribution $2k$ stems from the leading power $z^{2k}$ in the universal denominator \eqref{eq:denom_no_gauge} of $\Zmodk$ in the absence of divisors of type 3. The bound on the highest power in $E_2$ is thus
	$k+g-1$, which lies below the bound provided by the weight alone.

	The structure of $\cF_{\bk,g}$ in the presence of divisors of type 3 (i.e. of non-abelian gauge symmetry in the case of SCFTs) is similar. To arrive at this conclusion, consider again the general form
	\begin{equation}\label{eq:Z_jacobi_full}
	\Zmodk = \frac{\sum c_i \phi_{\bk,i}(\tau,z,\bm)}{\phi^D_{\bk}(\tau,z,\bm)} \,.
	\end{equation}
	of $\Zmodk$. The denominator, as given in
        \eqref{eq:denom-gauge}, now has $\bm$-dependence. Naively, the
        occurrence of linear combinations of $z$ and $\bm$ as elliptic
        parameters invalidates the above argument for the structure of
        $\cF_{\bk,g}$. Upon rewriting the denominator in the form
	\be \label{Z_jacobi_structure-2}
	\left(\prod_{s=1}^k \phi_{-2,1}(\tau,sz) \right) \prod_{\alpha \in \Delta_+} \phi_{-2,1}(\tau,m_\alpha)^{\alpha_k(0)} \left( \prod_{j=1}^{k-1} \prod_{\alpha \in \Delta_+}  \big( \phi_{-2,1}(\tau, j z + m_\alpha) \phi_{-2,1}(\tau, j z - m_\alpha) \big)^{\alpha_k(j)}\right) \,,
	\ee
	where
	\be \label{order_singular}
	\alpha_k(j) = \left \lceil \frac{k-j}{2} \right \rceil  \,,
	\ee
	we note however that each
	occurrence of the linear combination $jz + \ma$ as elliptic parameter of a $\phi_{-2,1}$ factor in $\phi^D_{\bk}$ is balanced by the occurrence of $jz-\ma$. The index bilinear form of the denominator thus disentangles the contribution of $z$ and of $\bm$ as elliptic parameters. The
	product over positive roots guarantees invariance under the action of
	the Weyl group. This is consistent with $\phi_{\bk}^D$ taking value in the tensor product of the rings of Jacobi forms with elliptic parameter $z$ and of Weyl invariant Jacobi forms with elliptic parameter $\bm$. Indeed, the identity
	\begin{equation}
	\phi_{-2,1}(z_1) \phi_{-2,1}(z_2) = \frac{1}{144}\left(\phi_{-2,1}(z_+)\phi_{0,1}(z_-) - \phi_{-2,1}(z_-)\phi_{0,1}(z_+)
	\right)^2 \,, \quad z_{\pm} = \tfrac{z_1\pm z_2}{2} \,,
	\end{equation}
	can be used to eliminate linear combinations of $z$ and $m_\alpha$ as elliptic arguments. The leading coefficient of $\phi^D_{\bk}(\tau,z,\bm)$ in $z$ is a Weyl invariant Jacobi form. Laurent expanding $\cF_{\bk}$ in $z$ thus yields $\cF_{\bk,g}$ as meromorphic Weyl invariant Jacobi forms in $\bm$ with coefficients in the ring of quasi-modular forms.
	
	As an example, we list low genus results for the free energy at base wrapping degree 2 on the geometry underlying the minimal 6d SCFT with gauge group $SU(3)$ (a vertical decompactification of the elliptic fibration over $\IF_3$), based on the calculation in \cite{DelZotto:2017mee}:
	\ba
	\cF_{2,0} &=& \frac{1}{\tpiI^2}\frac{5 \phi_{-2,3}^4 - 1792 \phi_{0,3} \phi_{-2,3} \phi_{-6,6} + 32 E_2 \phi_{-2,3}^2 \phi_{-6,6} + 
	512 E_4 \phi_{-6,6}^2}{4096 \phi_{-6,6}^3} \,, \nn\\
	\cF_{2,1} &=& \frac{1}{98304 \phi_{-6,6}^4}(3 \phi_{-2,3}^6 - 1568 \phi_{0,3} \phi_{-2,3}^3 \phi_{-6,6} + 50 E_2 \phi_{-2,3}^4 \phi_{-6,6} + 
	73728 \phi_{0,3}^2 \phi_{-6,6}^2 \nn\\&& - 17920 E_2 \phi_{0,3} \phi_{-2,3} \phi_{-6,6}^2 + 
	224 E_2^2 \phi_{-2,3}^2 \phi_{-6,6}^2 + 672 E_4 \phi_{-2,3}^2 \phi_{-6,6}^2 + 
	5120 E_2 E_4 \phi_{-6,6}^3 \nn\\&&+ 2048 E_6 \phi_{-6,6}^3) \,, \nn \\
	\cF_{2,2} &=& \frac{\tpiI^2}{754974720  \phi_{-6,6}^5}(495 \phi_{-2,3}^8 - 341760 \phi_{0,3} \phi_{-2,3}^5 \phi_{-6,6} + 9600 E_2 \phi_{-2,3}^6 \phi_{-6,6} \nn \\ &&+ 
	44892160 \phi_{0,3}^2 \phi_{-2,3}^2 \phi_{-6,6}^2 - 
	5017600 E_2 \phi_{0,3} \phi_{-2,3}^3 \phi_{-6,6}^2 + 80000 E_2^2 \phi_{-2,3}^4 \phi_{-6,6}^2 \nn \\ &&+ 
	149120 E_4 \phi_{-2,3}^4 \phi_{-6,6}^2 + 235929600 E_2 \phi_{0,3}^2 \phi_{-6,6}^3 - 
	28672000 E_2^2 \phi_{0,3} \phi_{-2,3} \phi_{-6,6}^3 \nn \\ &&- 
	22839296 E_4 \phi_{0,3} \phi_{-2,3} \phi_{-6,6}^3 + 266240 E_2^3 \phi_{-2,3}^2 \phi_{-6,6}^3 + 
	2158592 E_2 E_4 \phi_{-2,3}^2 \phi_{-6,6}^3 \nn \\&&+ 548864 E_6 \phi_{-2,3}^2 \phi_{-6,6}^3 + 
	8192000 E_2^2 E_4 \phi_{-6,6}^4 + 3342336 E_4^2 \phi_{-6,6}^4 + 
	6553600 E_2 E_6 \phi_{-6,6}^4) \nn \,.
	\ea
	This result is expressed in terms of the generators of the ring $J^D_{*,*}(\mathfrak{a}_2)$ introduced in appendix \ref{s:app_jac}. The map $\cF_{k,g} \rightarrow F_{k,g}$ is implemented by replacing these generators by their images in $J^{\widehat{D}}_{*,*}(\mathfrak{a}_2)$ under the map \eqref{eq:map_D_to_Dhat}, and dividing by $\eta(\tau)^{12k}$.

	\subsubsection{The refined case} \label{sss:transf_ref}
	In topological string applications, the refined free energy is sometimes usefully written as a function of the parameters $g_s^2 = - \epsilon_1 \epsilon_2$ and $s = \epsilon_1 + \epsilon_2$. The symmetry under exchange of $\epsilon_1$ and $\epsilon_2$, apparent in \eqref{GVref}, is manifest with this choice. For our purposes of retaining some of the transformation properties of the partition function in passing to the free energy, this choice at first glance appears propitious, as the index bilinear written in terms of these variables decomposes into a sum of squares, with the coefficient of $s^2$ depending linearly on $\bk$. This naively suggests that the expansion coefficients of $\cF_{\bk}$ in $g_s$ should have good transformation properties with regards to the remaining parameter $s$. This however is not the case, as unlike the superficially analogous case of $\bm$-dependence in the unrefined case, expansion in $g_s$ breaks the periodicity in the variable $s$ as well; $\cF_{\bk,g}(\tau,s)$ hence cannot be expanded in Jacobi forms in the elliptic parameter $s$. Expanding $\cF_{\bk}(\tau, z_L,z_R,\bm)$ therefore in both $z_L$ and $z_R$ (recall that $z_{\bullet} = \tfrac{\epsilon_{\bullet}}{2\pi}$), we obtain
	\be
	\cF_{\bk}(\tau,z_L,z_R,\bm) = \frac{1}{z_1 z_2}\sum_{g=0}^\infty \cF_{\bk,g_L,g_R}(\tau,\bm) \,z_L^{2g_L}z_R^{2g_R} \,.
	\ee
	
	The same arguments as in the unrefined case show that $\cF_{\bk,g_L,g_R}(\tau,\bm)$ are elements of the ring of Jacobi forms with elliptic index $\bm$, tensored over the ring of quasi-modular forms. The $\bm$-index is that of $\Zmodk$, and the weight increases with the indices $g_L$ and $g_R$.

\subsection{Pole structure of single-wrapping contributions to free energy} \label{s:polestructure1}

Divisors of type 3 modify the pole structure of $Z_{\bk}$, and therefore $f_{\bk}$ drastically, as can be seen in \eqref{eq:denom-gauge}. We will thus discuss geometries with and without such divisors separately in this section. 

\subsubsection{Absence of divisors of type 3 (no non-abelian gauge symmetry) -- unrefined}
\label{sss:no3unref}

We will argue that $f_{\bk}(z,\tau,\bm)$ as a function of $z = \tfrac{g_s}{2\pi}$ has the following pole structure:
\begin{enumerate}[label=\bfseries(\Roman*)] 
\item \label{p1} A second order pole at all integers, and no further
  poles on the real line. The Laurent coefficients at these poles are
  determined by the single-wrapping genus 0 Gromov-Witten invariants
  at base degree $\bk$.
\item \label{p2} All the other poles are at non-real $s$-torsion
  points for all integers $s \le \max\{k_i\}$. These poles are of
  order $2\ell$ and less, where $\ell = \sum_i \lfloor k_i/s\rfloor$. The Laurent coefficients of these poles are determined by the single-wrapping free energies $f_{\bk'}$, $|\bk'| < |\bk|$, together with the data specified in \ref{p1}.
\end{enumerate}

To show \ref{p1}, notice that the Gopakumar-Vafa presentation
\eqref{eq:fn} of the single-wrapping contributions in the unrefined
case becomes
\begin{equation} 
  f_{\bk}(z,\tau,\bm) = \sum_{g,d_\tau,\dm \geq 0}
  n_g^{\bk,d_\tau,\dm}\(2\sin \pi z\)^{2g-2}q^{d_\tau}
  \bQ_m^{\dm} \ .
\end{equation}
The second order poles at integer values of $z$ are visible from the
$\sin(\pi z)$ factor with $g=0$. Their Laurent coefficients depend
only on the single-wrapping genus 0 Gromov-Witten invariants; for example at $z\to 0$
\begin{equation}\label{eq:fpol-unref}
  f_{\bk}(z,\tau,\bm) = \frac{\sum_{d_\tau,\dm\geq 0}
    n_0^{\bk,d_\tau,\dm}q^{d_\tau}\bQ^{\dm}}{(2\pi)^2z^2} +
  \text{regular terms in }z \ .
\end{equation}
Note that we prefer speaking of single-wrapping genus 0 Gromov-Witten invariants, rather than the identical genus 0 Gopakumar-Vafa invariants, as we are concerned with the $z$ expansion of $f_{\bk}$, rather than the expansion in $\sin \pi z$.

Regarding the possibility of other real poles, the infinite sum in $q$ cannot lead to poles at $\tau$-independent (hence real) points, and the infinite sum in $\bQ_m$ does not lead to additional poles in the absence of divisors of type 3. $f_{\bk}$ hence has no
other poles on the real line.

To argue for \ref{p2}, consider the expression \eqref{eq:fZ-ref} of
the single-wrapping function $f_{\bk}(z,\bt)$ in terms of the
partition function $Z_{\bk'}$ at base wrappings $|\bk'| \leq |\bk|$.

$f_{\bk}(\tau,z,\bm)$ will at best share the poles of the base-wrapping $\bk'$ partition functions $Z_{\bk'}$ appearing on the RHS of \eqref{eq:fZ-ref}. These poles lie at $s$-torsion points for all integers $s\leq \max\{k_i\}$. The maximal order of a pole at an $s$-torsion point will arise from a contribution to \eqref{eq:fZ-ref} with a maximal number $\ell$ of $Z_{\bullet}$ factors at multi-wrapping $s$. This number is $\ell = \sum_i \lfloor k_i/s\rfloor$, and the order of the corresponding pole is $2 \ell$. 

By \ref{p1}, a certain number of cancellations must take place between the poles stemming from the monomials contributing to $f_{\bk}$:
\begin{itemize}
	\item All but the second order pole at integer values of $z$ cancel.
	\item All real poles at real $s$ torsion points, $s>1$, cancel.
\end{itemize}
We observe by explicit computation that generically, no other cancellations occur. For all non-real $s$-torsion points, $s>1$, this follows from the following observation: as we will see in detail in section \ref{s:computing_laurent_Z},  all poles at $s$-torsion points for a given $s$ are related via modular transformations. Given that the monomials contributing to $f_{\bk}$ have different indices, the vanishing of the poles at one element of this $SL(2,\IZ)$ orbit, the real $s$-torsion point, generically excludes the vanishing at all others.

We will show in section \ref{s:computing_laurent_Z} that the principal part of the partition function $Z_{\bk}$ at base-wrapping $\bk$ can be computed from knowledge of the partition function at lower base-wrapping, in conjunction with knowledge of the single-wrapping genus 0 Gromov-Witten invariants at base-wrapping $\bk$. The central ingredient in the argument is that all poles at $s$-torsion points for fixed $s$ are related by an $SL(2,\IZ)$ action, and each $SL(2,\IZ)$ orbit has a distinguished real representative; the principal part of the Laurent expansion of $Z_{\bk}$ around this point is determined by the data proposed. This property of $Z_{\bk}$ implies the final claim of \ref{p2} regarding the pole structure of $f_{\bk}(\tau,z,\bm)$.

\subsubsection{Absence of divisors of type 3 (no non-abelian gauge symmetry) -- refined}
	
As in the unrefined limit given by $\epsilon_1 = - \epsilon_2$, $\epsilon_L \propto g_s$ while $\epsilon_R = 0$, we will study the pole structure of $f_{\bk}(z_{L},z_{R},\bt)$ as a function of $z_L$. This choice is not necessarily canonical. Arguing in direct analogy to the unrefined case, we find the following:
\begin{enumerate}[label=\bfseries(\Roman*)]
\item\label{rp1} $f_{\bk}(z_{L},z_{R},\bt)$ as a function of
  $z_L$ has simple poles at
  $z_L = \pm z_R+ \,\IZ$ and no further poles along
  the axes $\pm z_R + \,\IR$. At these poles, the Laurent
  coefficients depend only on the refined GV invariants
  $n_g^{\bd,\text{NS}}$ appearing in the NS limit
  $\epsilon_1 \rightarrow 0$ (or equivalently
  $\epsilon_2 \rightarrow 0$) of the free energy, which are given by
  \begin{equation} \label{eq:NS-inv}
    n_g^{\bd,\text{NS}} = \sum_{g_L+g_R = g}n_{g_L,g_R}^{\bd} \ .
  \end{equation}
\item\label{rp2} There are further poles at $\pm z_R$ shifted by non-real
  $s$-torsion points for all integers $s\leq \max\{k_i\}$. These poles
  are of order $\ell = \sum_i \lfloor k_i/s \rfloor$ and less.
\end{enumerate}
The argument proceeds in precise analogy to the unrefined case in section \ref{sss:no3unref}. To see that the invariants \eqref{eq:NS-inv} play the role of the genus 0 Gromov-Witten invariants in the unrefined case, note that expanded around $z_L = \pm z_R$,
\begin{equation}\label{eq:fpol-ref}
\begin{aligned}
f_{\bk}(z_{L},z_R,\bt) =& \mp \frac{\sum_{d_\tau,\dm,g\geq 0}n_g^{n,d_\tau,\dm,\text{NS}}(2\sin\pi z_R)^{2g}/(2\sin 2\pi z_R) q^{d_\tau}\bQ_m^{\dm}}{ 2\pi (z_L \mp z_R)}
\\
&+ \text{regular terms in }{z_L \mp z_R}\ .
\end{aligned}
\end{equation}

\subsubsection{Presence of divisors of type 3 (gauge symmetry) - unrefined}

In geometries that exhibit divisors of type 3, the denominator $\phi_{\bk}^D$ exhibited in \eqref{eq:general_form} depends on the associated K\"ahler parameters, and $Z_{\bk}$ as a function of $z$ exhibits poles that depend on these parameters. These poles are inherited by the free energies $F_{\bk}$.

We will focus  on the case $b_2(B)=1$. From the explicit form of the denominator $\phi_{\bk}^D$ given in \eqref{Z_jacobi_structure-2}, we can read off the poles of the
partition function $Z_k$ at fixed base degree $k$:
\begin{enumerate}
\item Poles of order $2\lfloor k/s\rfloor$ at $s$-torsion points for
  $1 \leq s\leq k$.
\item Poles at
  \begin{equation}
    z_j = \frac{m_\alpha}{j} + j\text{-torsion point}
  \end{equation}
  for $1 \le j\leq k-1$ and $m_\alpha = (\bm,\alpha)$, $\alpha\in\Delta$, of order $2\alpha_k(j)$ as defined in \eqref{order_singular}.
\end{enumerate}
The poles of the first kind are independent of the presence of
divisors of type 3, and can be treated as in subsection
\ref{sss:no3unref}. The poles of the second kind are qualitatively
different, as no representative in their $SL(2,\IZ)$ orbit is related
to simply accessible data: they are all invisible in the Gopakumar-Vafa presentation \eqref{fgvtop} of the free energy. This limitation reduces the utility of analyzing the pole structure of $Z_k$ via knowledge of the pole structure of $f_k$. Having come this far, we nevertheless want to make some observations regarding the latter.

Let us first consider $F_k$ and the possibility that the poles of the monomials $M_{k,n}$ on the RHS of the expression \eqref{eq:FfromZ} for $F_k$ in terms of $Z_{k'}$, $k'\le k$ cancel. For $k-j$ odd, this is ruled out by the fact that $\alpha_k(j) > \alpha_{k'}(j)$ for $k' < k$. Hence, the order of the pole at $z_j$ is highest amongst the monomial $M_{k,n}$ in the one equaling $Z_k$, the pole at this order can therefore not be canceled. For $k-j$ even, $\alpha_k(j) = \alpha_{k-1}(j)$ and $\alpha_{k-1}(j) > \alpha_{k'-1}(j)$ for $k' < k$, so the $Z_k$ and the $ \propto Z_{k-1} Z_1$ contribution to $F_k$ have the same order pole at $z_j$. We have checked via explicit computation that these do not cancel. 

Turning now to $f_k$, by the multi-wrapping structure \eqref{eq:fn} of the free energy,
the single-wrapping quantities $f_k$ must exhibit poles beyond those
of $F_k$.  To see the origin of such poles, consider e.g. the base
degree $k=4$. We have
\begin{equation}
F_4(\tau,z,m_\alpha) = f_4(\tau,z,m_\alpha) +
\frac{1}{2}f_2(2\tau,2z, 2m_\alpha) + \frac{1}{4} f_1(4\tau,4z,
4m_\alpha) \,.
\end{equation}
As $f_2(\tau,z,m_\alpha)$ exhibits poles at $z= r + s \tau \pm \ma$,
$f_2(2\tau,2z,2m_\alpha)$ will exhibit poles at
$z= \frac{r}{2} + s \tau \pm \ma$. But as no the $Z_{\bullet}$ exhibits
poles at this location, neither can $F_4$. $f_1=Z_1$, hence does not exhibit a pole at this location either. The pole must hence be
canceled by $f_4$.

	\section{Computing the Laurent coefficients of $Z_{\bk}$} \label{s:computing_laurent_Z}
\subsection{Laurent coefficients of the unrefined partition	functions}

\label{sc:LC-unref}

Based on the relation \eqref{eq:rec-ref} relating $Z_{\bk}(\tau,z,\bm)$ to the single-wrapping free energy $f_{\bk}(\tau,z,\bm)$ and $Z_{\bk'}$, $|\bk'|< |\bk|$ and the properties of $f_{\bk}(\tau,z,\bm)$ that we have established in the previous section, we will now demonstrate that the principal parts of $Z_{\bk}(\tau,z,\bm)$ can be computed from the knowledge of $Z_{\bk'}$ with $|\bk'| < |\bk|$, complemented with the single-wrapping genus zero Gromov-Witten invariants at base degree $\bk$.

$Z_{\bk}$ has poles at all $s$-torsion points
\begin{equation}
z = \frac{c\tau+d}{s} 
\end{equation}
for $s \le \max\{k_i\}$. From property \ref{p1} of $f_{\bk}$
and \eqref{eq:rec-ref}, it follows that the knowledge of genus
zero Gromov-Witten invariants for base wrapping degree
$|\bk'| \le |\bk|$ is sufficient to fix the Laurent
coefficients of $Z_{\bk}$ at all real torsion points. Our task
is therefore to reduce the discussion at non-real torsion
points to this case. As the following discussion relies on the
$SL(2,\IZ)$ transformation behavior of $\Ztop$, it will be
convenient to work with the quantity $\Zmodk$, which differs
from $\Zk$ by a rescaling, see \eqref{eq:enum_to_mod}.

We consider hence a non-real $s$-torsion point
\begin{equation} \label{eq:nonrealtp}
z = r\frac{c\tau+d}{s} \ ,\quad c\neq 0,
\end{equation}
where, for $d \neq 0$, we enforce $(c,d)=1$ by including the factor $r$. We can relate the Laurent coefficients $\tc_{\bk,n}$ of an expansion around this point,
\begin{equation}\label{eq:Zkc}
\Zmodk(\tau,z,\bm) = \sum_{n=-2\ell}^\infty \tc_{\bk,n}(\tau,\bm)\(z-
r\frac{c\tau+d}{s}\)^n \ ,
\end{equation}
to those around the real torsion point $\tfrac{r}{s}$,
\begin{equation}
\Zmodk(\tau,z,\bm) = \sum_{n = - 2\ell}^\infty
\tb_{\bk,n}(\tau,\bm) (z - \tfrac{r}{s})^n \,.
\end{equation}
by considering the modular transformation
$\(\begin{smallmatrix}a & b \\ c & d \end{smallmatrix}\)\in
SL(2,\IZ)$ of this latter expression. We obtain
\begin{equation} \label{eq:matching_first_step}
\Phi_{w,\indexz(\bk),\indexm(\bk)}\Zmodk(\tau,z,\bm) =
\Zmodk(\tfrac{a \tau + b}{c \tau +d}, \tfrac{z}{c\tau+d},
\tfrac{\bm}{c\tau+d}) = \sum_{n = - 2\ell}^\infty
\frac{\tb_{\bk,n}(\tfrac{a\tau+b}{c\tau+d},\tfrac{\bm}{c\tau+d})}{(c\tau+d)^n}
\left(z - r\frac{c\tau+d}{s } \right)^n \,,
\end{equation}
where
\begin{equation}
\Phi_{w,\indexz(\bk),\indexm(\bk)} = (c\tau+d)^w \e\[\tfrac{\indexz(\bk)c
	z^2}{c\tau+d}\] \e\[\tfrac{\indexm(\bk)c (\bm,\bm)}{2(c\tau+d)}\] \,,
\end{equation}
with the notation $\e[x] = \exp(2\pi\ri x)$.
Note that unlike $\Zk$, $\Zmodk$ has non-trivial weight $w$. Substituting in the expression \eqref{eq:Zkc}
yields
\begin{multline} \label{eq:matching_LC}
\sum_{n = - 2\ell}^\infty
\tc_{\bk,n}(\tau,\bm) \(z - r\frac{c \tau+d}{s}\)^n = (c\tau+d)^{-w}
\e\[-\tfrac{\indexz(\bk)c z^2}{c\tau+d}\] \e\[-\tfrac{\indexm(\bk)c
	(\bm,\bm)}{2(c\tau+d)}\] \\ \sum_{n = - 2\ell}^\infty
\frac{\tb_{\bk,n}(\tfrac{a\tau+b}{c\tau+d},\tfrac{\bm}{c\tau+d})}{(c\tau+d)^n}
\(z - r\frac{c\tau+d}{s } \)^n \,.
\end{multline}
The Laurent coefficients $\tc_{\bk,n}$ of $\Zmodk$ around the non-real torsion point \eqref{eq:nonrealtp} can now be expressed  in terms of
the coefficients $\tb_{\bk,n}$ around the real torsion point $\tfrac{r}{s}$ by expanding the remaining $z$ dependence in the
exponential on the RHS of \eqref{eq:matching_LC}. 

Equation \eqref{eq:matching_LC} is difficult to use for explicit calculations away from $s=1$, as the expansion coefficients of Jacobi forms around $s$-torsion points do not behave well under modular transformations for $s>1$. For computations, we must obtain all expressions in an expansion in $q = \e[\tau]$. To put $\tb_{\bk,n}(\tfrac{a\tau+b}{c\tau+d},\tfrac{\bm}{c\tau+d})$ in this form, we proceed by induction. Assume that we have determined all $Z_{\bk'}$ for $|\bk'|<|\bk|$. From equation \eqref{eq:rec-ref} and property \ref{p1} of the single-wrapping free energy $f_{\bk}$, we see that the principal parts of $Z_{\bk}$ around real $s$-torsion points for $s > 1$ are determined completely by the monomials $\mki$ in $Z_{\bk'}$, $|\bk'|<|\bk|$ introduced in \eqref{eq:rec-ref_aux}, i.e. do not involve $f_{\bk}$. The negative index Laurent coefficients $\stb_{\bk,n}$ of $Z_{\bk}$ around the $s$-torsion point $\tfrac{r}{s}$, $s>1$, relate to those of each monomial $\mki$ around this point, which we denote as $\stb_{\bk,i,n}$, as
\be \label{eq:bs}
\stb_{\bk,n} = \sum_i \stb_{\bk,i,n} \,, \quad n<0 \,.
\ee
For each monomial, we introduce the product of Jacobi forms $\tMki$, which is $\mki$ evaluated on the set $\{ {\cal Z}_{\bk'} \}$ rather than $\{ {Z}_{\bk'} \}$, with the corresponding Laurent coefficients $\tb_{\bk,i,n}$, 
\be
\tMki(\tau,z,\bm) = \sum_{n=-2 \ell}^{\infty} \tb_{\bk,i,n}(\tau, \bm) \(z - \frac{r}{s}\)^n \,.
\ee
The coefficient $\tb_{\bk,i,n}$ evaluated at the point of interest satisfy
\be \label{eq:mk_transformed}
\tMki(\tfrac{a \tau + b}{c \tau +d}, \tfrac{z}{c\tau+d},
\tfrac{\bm}{c\tau+d}) = \sum_{n=-2 \ell}^{\infty} \frac{\tb_{\bk,i,n}(\tfrac{a\tau+b}{c\tau+d},\tfrac{\bm}{c\tau+d})}{(c\tau+d)^n}
\(z - r\frac{c\tau+d}{s } \)^n\,.
\ee
We cannot, in analogy with \eqref{eq:matching_first_step}, directly relate the LHS to $\tMki(\tau,z,\bm)$ via a modular transformation. This is because $\tMki$ will generically contain factors of Jacobi forms evaluated at arguments $(t \tau, t z , t\bm)$, $t > 1$, hence not be a Jacobi form for the full modular group $SL(2,\IZ)$. To obtain a $q$-expansion of $\tb_{\bk,i,n}(\tfrac{a\tau + b}{c\tau + d} \tfrac{\bm}{c \tau + d})$ nonetheless, we study the factors contributing to $\tMki(\tau,z,\bm)$ individually. A generic such factor is of the form ${\cal Z}_{\bk_{i,j}}(s_j \tau, s_j z, s_j \bm)$. We will write $t=s_j$ in the following argument to lighten the notation. On the LHS of \eqref{eq:mk_transformed}, this factor occurs evaluated at the arguments
\be \label{eq:Z_factor_in_monomial}
{\cal Z}_{\bk_{i,j}}(\t \tfrac{a \tau + b}{c \tau +d}, \t \tfrac{z}{c\tau+d}, \t	\tfrac{\bm}{c\tau+d}) \,.
\ee
Our goal is to obtain a form amenable to $q$-expansion of all such factors contributing to the monomial $\tMki(\tfrac{a \tau + b}{c \tau +d}, \tfrac{z}{c\tau+d},\tfrac{\bm}{c\tau+d})$. This will allow us to express the negative index coefficients on the RHS of \eqref{eq:mk_transformed} in terms of a $q$-expansion, which via \eqref{eq:bs} will yield the desired $q$-expansion of the coefficients $\tc_{\bk,n}(\tau,\bm)$ in \eqref{eq:matching_LC} .

The transformation
\be \label{eq:transf_not_sl2z}
\begin{pmatrix} d & - \t \,b \\ -c  & \t\, a  \end{pmatrix}  \, : \quad \t \,\frac{a \tau + b}{c \tau +d} \,\mapsto \,\tau
\ee
would remove the $\tau$-dependence of the denominator of the modular argument of \eqref{eq:Z_factor_in_monomial}, but is not an element of $SL(2,\IZ)$ for $\t \neq 1$. We can correct for this by adjusting the two top entries of the matrix, as only the bottom entries enter in removing the $\tau$-dependence in the denominator of $\t \,\tfrac{a \tau + b}{c \tau +d}$. For a solution to this problem to exist, the bottom entries must be mutually prime. Let therefore $u = \gcd(c,t)$. Then we can find $\alpha$ and $\beta$ such that 
\be \label{eq:the_sl}
\mathrm{SL}(2,\IZ) \ni \begin{pmatrix} \alpha & \beta \\ -\tfrac{c}{u}  & \tfrac{\t}{u}\, a  \end{pmatrix}  \, : \quad \t \,\frac{a \tau + b}{c \tau +d} \,\mapsto \, \frac{p \,\tau + q}{\t} \,, \quad  p, \, q \in \IZ \,,
\ee
with
\begin{equation}
p = u^2 \ ,\quad q = u(\alpha b t + \beta d) \ .
\end{equation}
Setting
\be \label{eq:primed_var}
\begin{pmatrix} \alpha & \beta \\ \gamma  & \delta  \end{pmatrix}  = 	\begin{pmatrix} \alpha & \beta \\ -\tfrac{c}{u}  & \tfrac{\t}{u}\, a  \end{pmatrix} \,, \quad (\tau', z', \bm') = (\t \tfrac{a \tau + b}{c \tau +d}, \t \tfrac{z}{c\tau+d}, \t	\tfrac{\bm}{c\tau+d}) \,,
\ee
we obtain 
\be\label{eq: m_final}
\Phi_{w,\indexz(\bk_{i,j}),\indexm(\bk_{i,j})}{\cal Z}_{\bk_{i,j}} (\tau', z', \bm') = {\cal Z}_{\bk_{i,j}} (\tfrac{\alpha \,\tau' + \beta}{\gamma \tau' + \delta},\tfrac{z'}{\gamma \tau'+\delta},\tfrac{\bm'}{\gamma \tau'+\delta}) = {\cal Z}_{\bk_{i,j}} (\tfrac{p \,\tau + q}{\t},u\,z,u \, \bm)\,,
\ee
where 
\begin{equation}
\Phi_{w,\indexz(\bk),\indexm(\bk)} = \left(\frac{t}{u(c \tau +d)}\right)^{w_{i,j}} \e\[-t\tfrac{\indexz(\bk_{i,j}) c z^2}{c\tau+d}\] \e\[-t\tfrac{\indexm(\bk_{i,j})c (\bm,\bm)}{2(c\tau+d)}\] \,.
\end{equation}

Recalling $t= s_j$ and the constraint \eqref{eq:constraint_sj}
on $s_j$, we note that by linearity of the index
$\indexm(\bk)$ in $\bk$, the $q$-expansion of the expansion
coefficients
$\tb_{\bk,i,n}(\tfrac{a\tau+b}{c\tau+d},\tfrac{\bm}{c\tau+d})$
of the monomials $\mki$ for all $i$ will exhibit the same
prefactor $\e\[\tfrac{\indexm(\bk) (\bm,\bm)}{2(c\tau+d)}\]$.
The expression for
$\tb_{\bk,n}(\tfrac{a\tau+b}{c\tau+d},\tfrac{\bm}{c\tau+d})$ ($n < 0$)
obtained from these by summing over $i$ therefore also
exhibits this prefactor. In determining the Laurent
coefficients $\tc_{\bk,n}$ via \eqref{eq:matching_LC}, this
prefactor will hence cancels against
$\e\[-\tfrac{\indexm(\bk)(\bm,\bm)}{2(c\tau+d)}\]$.

Note that the argument of the exponential in $z^2$ exhibits a $\tau$-dependent denominator. This cancels in the first few terms in the expansion of the exponential around $r \tfrac{c\tau+d}{s}$, but computations at higher base degree will require expanding beyond these terms. This $\tau$-dependence outside of the $q$-expansion must cancel at the end of the calculation of the negative index Laurent coefficients: as $\Zmodk$ is a Jacobi form, Laurent expansion around a torsion point cannot give rise to this type of dependence.  

\subsection{Example: the massive E-string}\label{sec: E string}
In this subsection, we will exemplify the discussion above on the massive E-string at base degree one and two. Its index bilinear can be read off from that of the massive refined higher rank E-strings given
 in appendix \ref{s:index_bilinear}. 

We introduce the following notation: $D_{z_0}\, f$ will stand for the principal part of $f(z)$ at $z = z_0$, and $\lc{z_0}{k}\, f$ for the coefficient of $(z - z_0)^{-k}$ in this expansion. Thus, if $f$ has a pole at $z_0$ of maximal order $\ell$, 
\be
D_{z_0}\, f = \sum_{n=-\ell}^{-1} a_n (z-z_0)^n  \,, \quad \lc{z_0}{k}\, f = a_k \,.
\ee
The genus zero Gromov-Witten invariants required as input for our calculation can be determined e.g. with the help of mirror symmetry. This approach naturally yields the full genus 0 free energy $\cF_{k,0}= \lc{0}{2} \cF_k$, even though the new data at each new base order is of course captured by the single-wrapping invariants, generated by $f_k$. We report here the results up to base degree 3:

\be \label{E1g0}
\lc{0}{2}\, {\cal F}_1 = - \tpiI^{-2} A_1 ,
\ee
\be \label{E2g0}
\lc{0}{2}\, {\cal F}_2 = \frac{\tpiI^{-2}}{96} (4 E_2 A_1^2 + 5 E_4 B_{2} + 3 E_6 A_2),
\ee
\be\label{E3g0}
\begin{split}
	\lc{0}{2}\, {\cal F}_3 = \frac{\tpiI^{-2}}{15552} \Big(9 A_1 \big[5B_2(3E_{2}E_{4} + 5 E_6) &+ 3A_2(5E_4^2 + 3E_2 E_6)\big] \\
	&+ 54 A_1^3 (E_2^2 - E_4) + 28 A_3 (E_4^3 - E_6^2)\Big).
\end{split}
\ee
Here $A_i,B_j$ are $E_8$ Weyl invariant Jacobi forms whose
explicit expressions
can be found for instance in the Appendix D of \cite{Gu:2017ccq}.
Note in particular the bound $k-1$ on the power of $E_2$ occurring in ${\cal F}_k$, as derived in section \ref{ss:transf_unref}.

For the base degree one partition function ${\cal Z}_1 $, the general ansatz \eqref{eq:general_form} together with the denominator \eqref{eq:denom_no_gauge} and the indices given in appendix \ref{s:index_bilinear} yield
\be
{\cal Z}_1 = \frac{\phi_1(\tau, z, \bm)}{\phi_{-2,1}(\tau,z)} \,.
\ee
Inside the fundamental parallelogram for $z$ spanned by 1 and $\tau$,
$\Zmodk$ only has a double pole at the origin, with coefficient
(\ref{E1g0}). The weight and $\bm$-index  of ${\cal Z}_1$, 6 and 1
respectively, fix the numerator $\phi_1(\tau, z, \bm)$ to have weight
4 and index 1. This uniquely fixes it to be \cite{Gu:2017ccq}
\be \label{eq:z1res}
{\cal Z}_1  =- \frac{A_1}{\phi_{-2,1}(\tau,z)} \,,
\ee
up to the proportionality factor $-1$, which can be determined from a single genus 0 Gromov-Witten invariant.

Next, for the case of base degree two, we have
\ba \label{z2fg}
{\cal Z}_2 (\tau,z,\bm)&=& {\cal F}_2 (\tau,z,\bm)+ \frac{1}{2} {\cal Z}_1^2(\tau,z,\bm)  \\\nn
&=&  {\tf}_2(\tau,z,\bm) + \frac{1}{2} \frac{\eta^{24}(\tau)}{\eta^{12}(2\tau)} {\cal Z}_1(2\tau, 2z,2\bm) + \frac{1}{2} {\cal Z}_1^2(\tau,z,\bm) \,,
\ea
i.e.
\be \label{eq:tms}
\tm_{2,1} = \frac{1}{2}  {\cal Z}_1(2\tau, 2z,2\bm) \,, \quad \tm_{2,2} =\frac{1}{2} {\cal Z}_1^2(\tau,z,\bm)  \,.
\ee

From the general form (\ref{eq:denom_no_gauge}) of the denominator of ${\cal Z}_k$, we can read off that within a fundamental parallelogram of $z$, ${\cal Z}_2$ exhibits poles at all 2-torsion points
\be \label{eq:poles_k2}
z=0 , \frac{1}{2}, \frac{\tau}{2}, \frac{\tau+1}{2} \,.
\ee
To find the corresponding principal parts, we follow the strategy explained in section \ref{sc:LC-unref}: we first determine the expansions around the real poles, and then relate the Laurent data at the non-real torsion points to these.

\paragraph{$\boldsymbol{z=0}$} The structure of the denominator (\ref{eq:denom_no_gauge}) when $k = 2$ tells us that there are a fourth and a second order pole at $z=0$. The fourth order pole is due to the contribution $\frac{1}{2} {\cal Z}_1^2$ in \eqref{z2fg}. Its Laurent coefficient can be determined as

\be
\lc{0}{4}\, {\cal Z}_2 = \frac{1}{2} \left(\lc{0}{2}\, {\cal Z}_1\right)^2 \,.
\ee
The same term also contributes to the second order pole, together with the genus 0 part of ${\cal F}_2$,
\be
\lc{0}{2}\, {\cal Z}_2 = (\lc{0}{2}\, {\cal Z}_1) (\lc{0}{0}\, {\cal Z}_1) + \lc{0}{2}\, {\cal F}_2 \,.
\ee

\paragraph{$\boldsymbol{z= \frac{1}{2}}$} By property \ref{p1} of $f_k$ and the presentation \eqref{eq:Zf-ref} of $Z_k$ in terms of $f_{\bullet}$, the second order pole at $z = \frac{1}{2}$ is due to the multi-covering contribution $\frac{1}{2} {\cal Z}_1 (2 \tau, 2z,2\bm)$ to ${\cal F}_2$,
\be
\lc{\frac{1}{2}}{2}\, {\cal Z}_2 = \frac{1}{2} \frac{\eta^{24}(\tau)}{\eta^{12}(2\tau)}\lc{\frac{1}{2}}{2}\, {\cal Z}_1(2\tau, 2z,2\bm) \,.
\ee

\paragraph{$\boldsymbol{z= \frac{\tau}{2}}$} The pole at $\frac{\tau}{2}$ is due to the contribution $\tf_2(\tau,z)$. To determine its Laurent data, we follow the reasoning explained in the previous subsection. The non-real torsion point $\frac{\tau}{2}$ is mapped to the real torsion point $\frac{1}{2}$ via the $SL(2,\IZ)$ matrix $\(\begin{smallmatrix}0 & -1 \\ 1 & 0 \end{smallmatrix}\)$. In terms of the parametrization in section \ref{sc:LC-ref}, we are considering a 2-torsion point, hence $s=2$, and $c=1$, $d=0$, $r=1$. 

We now compare the two Laurent expansions
\be \label{z2Lexps}
{\cal Z}_2 = \sum_{n=-2}^\infty \tc_{2,n}(\tau,\bm) (z- \tfrac{\tau}{2})^n = \sum_{n=-2}^\infty \tb_{2,n}(\tau,\bm) (z- \tfrac{1}{2})^n \,.
\ee

${\cal Z}_2$ has weight $w=12$ and indices $i_{z}(2) = - 3$, $i_{\bm}(2) =2$. Hence, specializing the equality (\ref{eq:matching_LC}), we obtain
\be
{\cal Z}_2(\tau,z) = \tau^{-w} \e\left[-\tfrac{ i_{z}(2)\, z^2}{\tau}\right] \e\left[-\tfrac{ i_{\bm}(2)\, (\bm,\bm)_{\mathfrak{e}_8}}{2\tau}\right] \sum_{n=-2}^\infty \frac{\tb_n ( - \tfrac{1}{\tau}, \tfrac{\bm}{\tau})}{\tau^{n}} (z - \tfrac{\tau}{2})^n \,.
\ee 
Comparing with  \eqref{z2Lexps} yields
\be\label{eq:basedeg2}
\sum_{n=-2}^\infty \tc_{2,n} ( \tau, \bm) ( z - \tfrac{\tau}{2})^n =  \tau^{-w} \e\left[-\tfrac{ i_{z}(2)\, z^2}{\tau}\right] \e\left[-\tfrac{ i_{\bm}(2)\, (\bm,\bm)_{\mathfrak{e}_8} }{2\tau}\right] \sum_{n=-2}^\infty \frac{\tb_{2,n} ( - \tfrac{1}{\tau}, \tfrac{\bm}{\tau})}{\tau^{n}} (z - \tfrac{\tau}{2})^n \,,
\ee
as a special case of our general formula (\ref{eq:matching_LC}).

Our next task is to rewrite $\tb_{2,n} ( - \tfrac{1}{\tau}, \tfrac{\bm}{\tau})$ for $n$ negative. Of the two monomials $\mathscr{m}_{2, i}(\tau, z, \bm)$ in \eqref{eq:tms}, only $\tm_{2,1}$ contributes to the poles at $\frac{1}{2}$. The discussion beginning below \eqref{eq:mk_transformed} thus applies with $t=2$, $u=1$.  We have
\be \label{eq:b_1_over_tau}
D_{\frac{1}{2}}\, \tm_{2,1}(\tau,z,\bm) = \sum_{n=-2}^{-1} \tb_{2,n,1}(\tau, \bm) (z - \frac{1}{2})^n \,,
\ee
with $\tb_{2,n}=\tb_{2,n,1}$ for $n<0$, such that
\be
\left( \frac{\eta^{24}(-\frac{1}{\tau})}{\eta^{12}(-\frac{2}{\tau})} \right) D_{ \frac{\tau}{2} }\tm_{2,1}(-\tfrac{1}{ \tau }, \tfrac{z}{\tau}, \tfrac{\bm}{\tau}) = \sum_{n=-2}^{-1} \frac{\tb_{2,n} ( - \tfrac{1}{\tau}, \tfrac{\bm}{\tau})}{\tau^{n}} (z - \tfrac{\tau}{2})^n,
\ee
with
\be \label{eq:m21notq}
\tm_{2,1}(-\tfrac{1}{ \tau }, \tfrac{z}{\tau}, \tfrac{\bm}{\tau})  = \frac{1}{2} {\cal Z}_{1}(-\tfrac{2}{ \tau }, \tfrac{2z}{\tau}, \tfrac{2\bm}{\tau}) \,,
\ee
i.e. we have $(\tau', z', \bm') = (-\tfrac{2}{\tau}, \tfrac{2z}{\tau}, \tfrac{2\bm}{\tau})$ in the notation of \eqref{eq:primed_var}. The matrix \eqref{eq:transf_not_sl2z}, which here specializes to $\(\begin{smallmatrix} 0 & -2 \\ 1  & 0  \end{smallmatrix} \)$, would map the modular argument $\tau' = -\tfrac{2}{\tau}$ on the RHS of \eqref{eq:m21notq} back to $\tau$, but it is clearly not an element of $SL(2,\IZ)$. Following the prescription above \eqref{eq:the_sl}, we replace this transformation by the element $\(\begin{smallmatrix} 0 & -1 \\ 1  & 0  \end{smallmatrix} \) \in SL(2,\IZ)$. Acting with this transformation on the RHS of \eqref{eq:m21notq}, we obtain 
\be
\tm_{2,1}(-\tfrac{1}{ \tau }, \tfrac{z}{\tau}, \tfrac{\bm}{\tau}) = \frac{1}{2}\left(\frac{\tau}{2}\right)^{w/2} \e\left[2\tfrac{ \indexz(1) z^2}{\tau}\right] \e\left[2\tfrac{\indexm(1)(\bm,\bm)_{\mathfrak{e}_8}}{2 \tau}\right]  {\cal Z}_1(\tfrac{\tau}{2}, z, \bm),
\ee
in accordance with \eqref{eq: m_final}. Using this result to obtain a $q$-expansion of the RHS of \eqref{eq:b_1_over_tau} and plugging into \eqref{eq:basedeg2} yields
\be
\begin{aligned}
	\sum_{n=-2}^{-1} &\tc_n ( \tau, \bm) ( z - \tfrac{\tau}{2})^n  =\\ 
	&=\ D_{ \frac{\tau}{2} } \tau^{-w} \e\left[-\tfrac{ i_{z}(2)\, z^2}{\tau}\right] \e\left[-\tfrac{ i_{\bm}(2)\, (\bm,\bm)_{\mathfrak{e}_8} }{2\tau}\right] \left( \frac{\eta^{24}(-\frac{1}{\tau})}{\eta^{12}(-\frac{2}{\tau})} \right) \tm_{2,1}(-\tfrac{1}{ \tau }, \tfrac{z}{\tau}, \tfrac{\bm}{\tau})\\
	&=\ \ \frac{1}{2} D_{\frac{\tau}{2}} \left(\frac{1}{2\tau}\right)^{w/2} \e\left[-\tfrac{\indexz(2)  -2 \indexz(1)}{\tau}z^2\right] \left( \frac{\eta^{24}(-\frac{1}{\tau})}{\eta^{12}(-\frac{2}{\tau})} \right) {\cal Z}_1(\tfrac{\tau}{2}, z, \bm)\\
	&= -\frac{1}{2} D_{\frac{\tau}{2}} \e\left[-\tfrac{\indexz(2)  -2 \indexz(1)}{\tau}z^2\right] \left( \frac{\eta^{24}(\tau)}{\eta^{12}(\frac{\tau}{2})} \right) {\cal Z}_1(\tfrac{\tau}{2}, z, \bm).
\end{aligned}
\ee
The overall minus sign is due to the non-trivial multiplier system of the Dedekind $\eta$ function. Note the announced cancellation of the exponential prefactor in $(\bm,\bm)_{\mathfrak{e}_8}$. Expanding the RHS around $z= \frac{\tau}{2}$ and comparing coefficients finally yields
\be
\begin{aligned}
\lc{\frac{\tau}{2}}{2}\, {\cal Z}_2 (\tau,\bm) &= \tc_{-2}(\tau,\bm) = -\frac{1}{2}q^{\tfrac{1}{4}} \left( \frac{\eta^{24}(\tau)}{\eta^{12}(\frac{\tau}{2})} \right) \lc{\frac{\tau}{2}}{2} {\cal Z}_1(\tfrac{\tau}{2}, z, \bm) \,, \\
\lc{\frac{\tau}{2}}{1}\, {\cal Z}_2(\tau,\bm) &=  \tc_{-1}(\tau,\bm) = -\tpiI i_{z}(2) \tc_{-2}(\tau,\bm) \,,
\end{aligned}
\ee
where we have evaluated $\indexz(2)-2\indexz(1)=-1$, and used the fact that $\cZ_1$ only has a second order pole at $z=\tfrac{\tau}{2}$, as can be seen from its explicit form \eqref{eq:z1res}.

\paragraph{$\boldsymbol{z= \tfrac{\tau+1}{2}}$} We can determine the Laurent coefficients at $z= \tfrac{\tau+1}{2}$ by invoking the periodicity of ${\cal Z}_2$ in $\tau$:
\ba
{\cal Z}_2(\tau,z,\bm) &=& \sum_{n=-2}^\infty \tc_{n}(\tau,\bm) (z - \tfrac{\tau}{2})^n = \sum_{n=-2}^\infty \td_{n}(\tau,,\bm) (z - \tfrac{\tau+1}{2})^n \nn\\
&=& {\cal Z}_2(\tau+1,z,\bm) = \sum_{n=-2}^\infty \tc_{n}(\tau+1,\bm) (z - \tfrac{\tau+1}{2})^n \,.
\ea
Hence
\be 
\td_n(\tau,\bm) = \tc_n(\tau + 1,\bm) \,,
\ee
yielding
\be
\lc{\frac{\tau+1}{2}}{2}\, {\cal Z}_2 = \lc{\frac{\tau}{2}}{2}\, {\cal Z}_2(\tau + 1, z, \bm)  \,, 
\ee 
and
\be
\lc{\frac{\tau+1}{2}}{1}\, {\cal Z}_2 = \lc{\frac{\tau}{2}}{1}\, {\cal Z}_2(\tau + 1, z, \bm) \,.
\ee
We have thus expressed the principal parts of ${\cal Z}_2$ around all of its poles \eqref{eq:poles_k2} using the knowledge of ${\cal Z}_1$ and of genus zero Gromov-Witten data at base degree two.

Below, we list explicitly the Laurent data of base degree one and two: 
\be \nn
\lc{0}{2}\, {\cal Z}_1 = - \tpiI^{-2} A_1 \,.
\ee
\be \nn
\lc{0}{4}\, {\cal Z}_2 = \frac{\tpiI^{-4}}{2} A_1^2 \,.
\ee
\be \nn
\lc{0}{2}\, {\cal Z}_2 =  - \frac{\tpiI^{-2}}{96} \left( 12 A_1^2 E_2 + 5 B_2 E_4 + 3 A_2 E_6 \right) \,,
\ee 
\be \nn
\lc{\frac{1}{2}}{2}\, {\cal Z}_2 = -\frac{\tpiI^{-2}}{8} \left(\frac{\eta^{24}(\tau)}{\eta^{12}(2\tau)} \right) A_1(2\tau, 2\bm) \,,
\ee
\be \nn
\lc{\frac{\tau}{2}}{2}\, {\cal Z}_2 = \frac{\tpiI^{-2}}{2} \left( \frac{\eta^{24}(\tau)}{\eta^{12}(\tfrac{\tau}{2})} \right) q^{3/4} A_1(\tfrac{\tau}{2}, \bm) \,, 
\ee 
\be \nn
\lc{\frac{\tau}{2}}{1}\, {\cal Z}_2 = 3\frac{\tpiI^{-1}}{2} \left( \frac{\eta^{24}(\tau)}{\eta^{12}(\tfrac{\tau}{2})} \right) q^{3/4} A_1(\tfrac{\tau}{2}, \bm) \,,
\ee
\be \nn
\lc{\frac{\tau+1}{2}}{2}\, {\cal Z}_2 = -\ri \frac{\tpiI^{-2}}{2} \left( \frac{\eta^{24}(\tau)}{\eta^{12}(\tfrac{\tau+1}{2})} \right) q^{3/4} A_1(\tfrac{\tau + 1}{2}, \bm) \,, 
\ee 
and
\be \nn
\lc{\frac{\tau+1}{2}}{1}\, {\cal Z}_2 = -3 \ri \frac{\tpiI^{-1}}{2} \left( \frac{\eta^{24}(\tau)}{\eta^{12}(\tfrac{\tau+1}{2})} \right) q^{3/4} A_1(\tfrac{\tau + 1}{2}, \bm) \,.
\ee
We extend these calculations to base degree 3, restricting to the massless case for simplicity, in appendix \ref{sec: deg3}.

	\subsection{Laurent coefficients of the refined partition functions}
	
	\label{sc:LC-ref}
	
	The discussion for the refined partition function follows the
        same pattern as in the unrefined case; the additional data
        upon increasing the base-wrapping degree to $\bk$ are now the
        NS-invariants \eqref{eq:NS-inv} at base degree $\bk$.

        $Z_{\bk}$ as a function of $z_L$ has poles at $\pm z_R$
        shifted by all $s$-torsion points
        \begin{equation}
          z_L = \pm z_R + \frac{c\tau+d}{s}
        \end{equation}
        for $s\le \max\{k_i\}$. Due to the property \ref{rp1} of
        single wrapping contributions $f_{\bk}$ to the refined free
        energy, and the relation \eqref{eq:rec-ref} adapted by
        replacing $z$ everywhere by $z_L,z_R$, the Laurent
        coefficients of $Z_{\bk}$ at all the poles
        $z_L = \pm z_R + \text{real torsion}$ can be fixed by the
        knowledge of the NS-invariants at base wrapping degree
        $|\bk'| \le |\bk|$. Poles at $\pm z_R$ shifted by non-real
        torsion points, on the other hand, can be related to the former by a modular transformation. As in the unrefined
        case, we will for the rest of the discussion switch to the quantity $\Zmodk$,
        related to $\Zk$ by the rescaling given in
        \eqref{eq:enum_to_mod}, as $\Zmodk$ has better modular
        transformation properties.

        The Laurent coefficients $\tc_{\bk,n}(\tau,z_R,\bm)$ of an
        expansion around a nonreal-torsion-shifted singular point,
        \begin{equation}\label{eq:Zkc-ref}
          \Zmodk(\tau,z_L,z_R,\bm)
          = \sum_{n=-\ell}^\infty
          \tc_{\bk,n}(\tau,z_R,\bm)\(z_L\mp z_R- 
          r\frac{c\tau+d}{s}\)^n \ ,
	\end{equation}
        can be expressed in terms of the coefficients
        $\tb_{\bk,n}(\tau,z_R,\bm)$ around the real-torsion-shifted
        point $\pm z_R +\tfrac{r}{s}$
        \begin{equation}
          \Zmodk(\tau,z_L,z_R,\bm)
          = \sum_{n = - \ell}^\infty
          \tb_{\bk,n}(\tau,z_R,\bm) \(z_L\mp z_R - \frac{r}{s}\)^n \,.
	\end{equation}
        by expanding the following identity
	\begin{multline} \label{eq:matching_LC_ref}
	\sum_{n = - \ell}^\infty
	\tc_{\bk,n}(\tau,z_R,\bm)
        \(\hat{z}_L  - \, r\frac{c \tau+d}{s}\)^n = (c\tau+d)^{-w}
	\e\[-\tfrac{\indexL(\bk) c
            (\hat{z}_L^2 \pm 2 z_R\hat{z}_L)}{c\tau+d}\]
        \e\[-\tfrac{\index1(\bk)c z_R^2}{c\tau+d}\] \\ \e\[-\tfrac{\indexm(\bk)
		c (\bm,\bm)}{2(c\tau+d)}\]\sum_{n = - \ell}^\infty
            \frac{\tb_{\bk,n}(\tfrac{a\tau+b}{c\tau+d},\tfrac{z_R}{c\tau+d},
              \tfrac{\bm}{c\tau+d})}{(c\tau+d)^n}
	\(\hat{z}_L  - \,  r\frac{c\tau+d}{s } \)^n \,.
      \end{multline}
      Here we have introduced $\index1(\bk) = \indexL(\bk) + \indexR(\bk)$
      and the variable
      \begin{equation}
        \hat{z}_L = z_L \mp z_R \ ,
      \end{equation}
      evaluated in the neighborhood of $r(c\tau+d)/s$. As
      in the unrefined case, we also enforce $\text{gcd}(c,d)=1$ by
      introducing $r$.  Beside, we need to put
      $\tb_{\bk,n}(\tfrac{a\tau+b}{c\tau+d},\tfrac{z_R}{c\tau+d},
      \tfrac{\bm}{c\tau+d})$ in a form amenable to expansion in
      $q = \e[\tau]$.  We first note that the coefficients
      $\stb_{\bk,n}$ are related to those of the monomials $m_{\bk,i}$
      introduced in \eqref{eq:rec-ref} at the same torsion point by
      \begin{equation}
        \stb_{\bk,n} = \sum_i \stb_{\bk,i,n} \,, \quad n<0\ .
      \end{equation}      
      The monomials $\tm_{\bk,i}$, which are $m_{\bk,i}$ evaluated
      on the set $\{ {\cal Z}_{\bk'} \}$ rather than
      $\{ {Z}_{\bk'} \}$, are products of Jacobi forms. Their expansion coefficients $\tb_{\bk,i,n}$ around $\hat{z}= \frac{r}{s}$, evaluated at the point of interest,
      satisfy
      \begin{equation}
	\tMki(\tfrac{a \tau + b}{c \tau +d},
        \tfrac{z_L}{c\tau+d},\tfrac{z_R}{c\tau+d},
        \tfrac{\bm}{c\tau+d})
        = \sum_{n=-\ell}^\infty
        \frac{\tb_{\bk,i,n}(\tfrac{a\tau+b}{c\tau+d},\tfrac{z_R}{c\tau+d},
          \tfrac{\bm}{c\tau+d})}{(c\tau+d)^n}
        \(\hat{z}_L- r\frac{c\tau+d}{s } \)^n\,.
      \end{equation}
      To obtain a $q$-expansion of the negative order coefficients, we need to
      consider the individual factors contributing to the monomial  $\tm_{\bk,i}$, which are of the form
      \begin{equation}
        {\cal Z}_{\bk_i,j}
        (s_j\tfrac{a\tau+b}{c\tau+d},s_j\tfrac{z_L}{c\tau+d},
        s_j\tfrac{z_R}{c\tau+d},
        s_j\tfrac{\bm}{c\tau+d}) \ .
      \end{equation}
      with possible multi-wrapping factors $s_j$. With 
      \begin{equation}
	\begin{pmatrix} \alpha & \beta \\ \gamma  &
          \delta  \end{pmatrix}  = 	\begin{pmatrix} \alpha & \beta
          \\ -\tfrac{c}{u}  & \tfrac{\t}{u}\, a  \end{pmatrix} \,,
        \quad (\tau', z'_L, z'_R, \bm') = (s_j \tfrac{a \tau + b}{c
          \tau +d}, s_j \tfrac{z_L}{c\tau+d}, s_j
        \tfrac{z_R}{c\tau+d}, s_j \tfrac{\bm}{c\tau+d}) \,,
      \end{equation}
      and $u=\text{gcd}(c,t)$, 
      we find
      \begin{align}
	\Phi_{w,\indexL(\bk_{i,j}),\indexR(\bk_{i,j}),\indexm(\bk_{i,j})}{\cal
        Z}_{\bk_{i,j}} (\tau', z'_L,z'_R, \bm')
        &= {\cal Z}_{\bk_{i,j}}
        (\tfrac{\alpha \,\tau' + \beta}{\gamma \tau' +
          \delta},\tfrac{z'_L}{\gamma \tau'+\delta},
          \tfrac{z'_R}{\gamma \tau'+\delta},\tfrac{\bm'}{\gamma
          \tau'+\delta}) \nn\\
        &= {\cal Z}_{\bk_{i,j}} (\tfrac{p \,\tau +
          q}{s_j},uz_L,uz_R,u \, \bm)\,, \label{eq: m_final_ref}
      \end{align}
      where
      \begin{equation}
        p = u^2 \ ,\quad q = u(\alpha b s_j + \beta d) \ ,
      \end{equation}      
      and
      \begin{equation}
	\Phi_{w,\indexL(\bk),\indexR(\bk),\indexm(\bk)}
        =  \left(\frac{s_j}{u(c \tau +d)}\right)^w
        \e\[-s_j\tfrac{\indexL(\bk_{i,j}) c (\hat{z}_L^2\pm 2z_R
            \hat{z}_L)}
          {c\tau+d}\]
        \e\[-s_j\tfrac{\index1(\bk_{i,j})c z_R^2}{(c\tau+d)}\]
        \e\[-s_j\tfrac{\indexm(\bk_{i,j})c (\bm,\bm)}{2(c\tau+d)}\] \,.
      \end{equation}
      We note that both $\index1(\bk)$ and $\indexm(\bk)$ are linear in
      $\bk$. As a result, the expansion coefficients
      $\tb_{\bk,i,n}(\tfrac{a\tau+b}{c\tau+d}, \tfrac{z_R}{c\tau+d},
      \tfrac{\bm}{c\tau+d})$ of the monomials $\tm_{\bk,i}$ for all
      $i$ and thus their sum
      $\tb_{\bk,n}(\tfrac{a\tau+b}{c\tau+d}, \tfrac{z_R}{c\tau+d},
      \tfrac{\bm}{c\tau+d})$ will exhibit the same prefactors
      $\e\[\tfrac{\index1(\bk)c z_R^2}{(c\tau+d)}\]$,
      $\e\[\tfrac{\indexm(\bk)c (\bm,\bm)}{2(c\tau+d)}\]$, which then
      cancel against
      $\e\[-\tfrac{\index1(\bk)cz_R^2}{(c\tau+d)}\]$,
      $\e\[-\tfrac{\indexm(\bk)c(\bm,\bm)}{2(c\tau+d)}\]$ in
      \eqref{eq:matching_LC_ref} when one computes the Laurent
      coefficients $\tc_{\bk,n}$.

	\section{$Z_{\bk}$ of negative index} \label{s:nidx}
        
	Above, we have shown that in the case of elliptically fibered Calabi-Yau manifolds without divisors of type 3, the negative index Laurent coefficients of $Z_{\bk}$ at all poles can be reconstructed from the knowledge of $Z_{\bk'}$, $|\bk'|<|\bk|$, complemented by the genus 0 Gromov-Witten at base degree $\bk$. By a general property of Jacobi forms, when the $z$-index of $Z_{\bk}$ is negative, the knowledge of these coefficients is sufficient to reconstruct all of $Z_{\bk}$ \cite{Bringmann2016,Haghighat:2015ega}.
	
	Rather than relying on the pole structure, the input used in \cite{Gu:2017ccq,DelZotto:2017mee} to determine $Z_{\bk}$ are vanishing conditions on Gopakumar-Vafa invariants. On elliptic fibrations without divisors of type 3, these conditions suffice to determine $Z_{\bk}$ completely when its $z$-index shifted by +1 is negative \cite{Gu:2017ccq}. This is not the case in the presence of such divisors \cite{DelZotto:2017mee}.
	
	In the following subsection, we explain why the negativity of the index enters in both approaches. We then review the technique presented in \cite{Bringmann2016} to explicitly construct a negative index Jacobi form from its principal part. We finally discuss which class of models this technique can be applied to, arriving at the class of E-strings which features in the title of this paper.

	\subsection{Why negative index is simpler}
	The two observations that
	\begin{itemize}
		\item negative index Jacobi forms are determined by their principal parts, and 
		\item imposing generic vanishing conditions on Gopakumar-Vafa invariants suffices to determine $Z_{\bk}$ on elliptic Calabi-Yau manifolds in the absence of type 3 divisors when the $z$-index of $Z_{\bk}$ shifted by +1 is negative 
	\end{itemize}
	hold because holomorphic Jacobi forms of negative index do not exist.
	
	Indeed, for the first statement, consider two meromorphic Jacobi forms of equal weight, index, and principal parts. Their difference is again a Jacobi form of same weight and index as before, but with vanishing principle parts. If the index is negative, this difference, by the non-existence of non-trivial holomorphic Jacobi forms of negative index, is hence zero. The statement follows.
	
	Combined with our demonstration in this paper that in the absence of divisors of type 3, the principal parts of $\Zk$ can be determined from the knowledge of $Z_{\bk'}$, $|\bk'| < |\bk|$, and genus 0 Gromov-Witten invariants at base-wrapping degree $\bk$, this allows for the computation of $\Zk$ in geometries which satisfy the negativity constraint on the $z$-index.
	
	Before we come to the proof of the second statement, let us
        briefly review the necessary background. The Gopakumar-Vafa
        invariants $n_g^{\bd}$ as introduced in \eqref{fgvtop}  vanish
        at fixed $\bd$ for sufficiently high $g$, reflecting a bound
        on the maximal genus of a degree $\bd$ curve. Now consider the
        single-wrapping free energy \eqref{eq:fn} at base wrapping
        $\bk$, which by \eqref{eq:fZ-ref} satisfies 
	\be
	 f_{\bk}(z,\bt) = Z_{\bk}(z,\bt) +
	\pZf(Z_{|\bk'|<|\bk|}(*z,*\bt)) \ .
	\ee 
	Having computed $Z_{\bk'}$ for $|\bk'|< |\bk|$,  the invariants $n_g^{\bk,d_\tau,d_{\bm}}$ determining $f_{\bk}$ will depend linearly on the coefficients $c_i$ of the general ansatz \eqref{eq:general_form} for $Z_{\bk}$. By the sufficiency of generic vanishing conditions to solve for $\Zk$, we mean that imposing that $n_g^{\bk,d_\tau,d_{\bm}}= 0$ for $g \gg 0$, without the need to specify the bound on $g$ beyond which this vanishing occurs, suffices to uniquely determine the $c_i$. Now we come to the proof of the second statement. Assume the $Z_{\bk'}$ for all $|\bk'|< |\bk|$ have been determined. Let $Z^1_{\bk}$ and $Z^2_{\bk}$ be two possible extensions to base wrapping $\bk$, both satisfying the vanishing conditions. We have
	\be
	f_{\bk}^1 - f_{\bk}^2 = Z^1_{\bk} - Z^2_{\bk} \,,
	\ee
	as all multi-wrapping contributions (stemming from $|\bk'|< |\bk|$) cancel. The RHS of this equality tells us that the difference is a Jacobi form of same weight and index as $Z^{1,2}_{\bk}$, while the LHS, by the Gopakumar-Vafa form \eqref{eq:fn} of the free energy, implies that the only real poles of this difference are of second order and lie at integer $z$. This allows us to write\footnote{This point is argued for more carefully in \cite{Gu:2017ccq}.}
	\be \label{eq:diff_smooth}
	Z^1_{\bk} - Z^2_{\bk} = \frac{\phi_{\bk}(\tau,z,\bm)}{\phi_{-2,1}(\tau,z)} \,.
	\ee
	All other terms in the denominator \eqref{eq:denom_no_gauge} must cancel against the numerator. $\phi_{\bk}$ is a holomorphic Jacobi form of index one higher than that of $Z^{1,2}_{\bk}$. If the latter index +1  is negative, $\phi_{\bk}$ vanishes, and the second statement follows.
	
	Let us also point out why the argument fails for elliptic fibrations in the presence of type 3 curves. The denominator in this case is given by \eqref{Z_jacobi_structure-2}. The mass dependent poles are not visible in the Gopakumar-Vafa form, and as we say in section \ref{s:structureF}, they do not distribute nicely with regard to multi-wrapping considerations. Hence, the difference \eqref{eq:diff_smooth} in the singular case takes the more complicated form
	\begin{multline} \label{eq:diff_smooth}
	Z^1_{k} - Z^2_{k} =\\ \frac{\phi_{\bk}(\tau,z,\bm)}{\phi_{-2,1}(\tau,z) \prod_{\alpha \in \Delta_+} \phi_{-2,1}(\tau,m_\alpha)^{\alpha_k(0)} \left( \prod_{j=1}^{k-1} \prod_{\alpha \in \Delta_+}  \big( \phi_{-2,1}(\tau, j z + m_\alpha) \phi_{-2,1}(\tau, j z - m_\alpha) \big)^{\alpha_k(j)}\right) } \,.
	\end{multline}
	To argue for vanishing as above, the sum of the index of the LHS and the index of the denominator of the RHS must be negative, a much weaker constraint than previously.
	
	\subsection{Explicit reconstruction of Jacobi forms from their principal parts} \label{s:explicit_reconstruction}
	The construction presented in \cite{Bringmann2016} relies on the existence of a function $F_M(\tau,z,u)$ for any $M \in \frac{1}{2} \IN$, which satisfies the following two properties:
	\begin{enumerate}
		\item It is quasi-periodic as a function of $u$, i.e. it satisfies the relation
		\be
		F_M(\tau,z,u+\lambda \tau + \mu) = (-1)^{2M \mu} \e\[-M(\lambda^2\tau + 2 \lambda u)\] F_M(\tau,z,u) \,.
		\ee
		For $M \in \IN$, this is the elliptic transformation property of a Jacobi form of index $M$.
		\item Again as a function of $u$, it is meromorphic with only simple poles. These lie at $z + \IZ \tau + \IZ$. The residue at $u=z$ is $\frac{1}{2\pi \ri}$.
	\end{enumerate}
	Now consider a Jacobi form $\phi_{-M}(\tau,u)$ of index $-M$, $M \in \IN$. The product
	\be \label{eq:phiF}
	\phi_{-M}(\tau,u) F_M(\tau,z,u)
	\ee
	is one- and $\tau$-periodic in the variable $u$. The integral around the boundary of a fundamental parallelogram for the lattice $\IZ \tau + \IZ$, chosen to avoid all poles of the integrand \eqref{eq:phiF} along the integration path, thus vanishes,
	\be \label{eq:vanint}
	\inttorus \phi_{-M}(\tau,u) F_M(\tau,z,u)\,du =0 \,.
	\ee
	Now choose $z$ away from the poles of $\phi_{-M}(\tau,u)$. Evaluating the LHS of \eqref{eq:vanint} by the residue theorem yields
	\be \label{eq:phiRes}
	\phi_{-M}(\tau,z) = - 2\pi \ri \sum_i \Res_{u=u_i} ( \phi_{-M}(\tau,u)F_M(\tau,z,u)) \,,
	\ee
	with the sum ranging over all poles of $\phi_{-M}$ in the chosen fundamental domain. This is the desired result, as the RHS of this expression is completely determined by $F_M$ and the negative index Laurent coefficients at the poles of $\phi_{-M}$.
	
	Explicitly, $F_M$ is a level 2M Appell-Lerch sum, given by the expression
	\be
	F_M(\tau,z,u) = (\zeta \omega^{-1}) \sum_{n \in \IZ} \frac{\omega^{-2Mn} q^{Mn(n+1)}}{1-q^n \zeta \omega^{-1}} \,, \quad q=\e[\tau], \, \omega=\e[u] , \, \zeta = \e[z]\,.
	\ee
	In terms of the principal parts $\LC_{n,u_i}$ of $\phi_{-M}(\tau,z)$ at the poles $u_i$,
	\be
	\phi_{-M}(\tau,z) = \sum_{n<0} \LC_{n,u_i} (z-u_i)^{n} + \mbox{holomorphic} \,,
	\ee
	the RHS of \eqref{eq:phiRes} can be evaluated to
	\be
	\phi_{-M}(\tau,z) = \sum_i \sum_{n>0} \frac{\LC_{-n,u_i}}{(n-1)!} \left[\left(\frac{1}{2\pi \ri}\frac{\partial}{\partial u}\right)^{n-1} F_{M}(\tau,z,u)\right]_{u=u_i} \,.
	\ee

We have used this procedure as a check on our computation in section \ref{sec: E string} and in appendix \ref{sec: deg3} of the principal parts of $Z_k$ for the massive E-string at $k=2$ and the massless E-string at $k=3$ respectively.

\subsection{Geometries on which $\Zk$ is completely determined by genus 0 Gromov-Witten invariants}
The $z$-index of $\Zk$ is given by the formula
\be
\indexz(\bk) = \frac{C_{\bk} \cdot (C_{\bk} + K_B)}{2} \,,
\ee
with the $C_{\bk}$ denoting divisors of the base $B$ of the elliptically fibered Calabi-Yau manifold $X$. For this to be negative for all $\bk$, we need $C_{\bk} \cdot C_{\bk} < 0$, as the second contribution to $\indexz(k)$ grows only linearly in $\bk$. In particular, we have to exclude base surfaces with divisors of positive self-intersection number, hence all compact projective surfaces. Luckily, we are often interested in considering non-compact base surfaces, e.g. when engineering 6d SCFTs in F-theory. 

To allow for the determination of the principal parts of $\Zk$ from genus 0 Gromov-Witten invariants, we need to exclude divisors of type 3 in the Calabi-Yau manifold $X$. This leaves us with geometries containing only curves of self-intersection number $-2$ or $-1$. By the analysis of \cite{Heckman:2015bfa}, the most general such configuration leading to a minimal SCFT is a chain of $-2$ curves ending with a $-1$ curve, with neighboring curves intersecting once. This is the higher rank E-string. Its $z$-index is given in \cite{Gu:2017ccq}. Considering a chain of length $n+1$, with the $0^{th}$ node indicating the $-1$ curve, and denoting the corresponding intersection matrix as $C_{IJ}$, the index can be written as
\be
\indexz(\bk) = \frac{1}{2} \( \sum_{I,J=0}^n k_I C_{IJ} k_J - k_0\) \,,
\ee
which is clearly negative definite.

\section*{Acknowledgements}
We are grateful to Albrecht Klemm, Guglielmo Lockhart, Andre Lukas, and
David Morrison for discussions. We acknowledge support from the grant
ANR-13-BS05-0001. JG is supported by the European Research Council
(Programme ``Ideas'' ERC-2012-AdG 320769 AdS-CFT-solvable).

 \newpage
 
  \appendix
  
  \section{Jacobi forms: the rings $J$, $J(\mathfrak{g})$, $J^D(\mathfrak{g})$, and $J^{\widehat{D}}(\mathfrak{g})$ } \label{s:app_jac}
	
	Jacobi forms~\cite{EZ} are holomorphic functions $\phi_{k,m}:\mathbb{H}\times \mathbb{C}^n\rightarrow \mathbb{C}^n$ that depend on a modular 
	parameter $\tau\in \mathbb{H}$ and an elliptic parameters $z\in \mathbb{C}$. They satisfy the transformation equation
	\be
	\phi_{k,m}\left(\tfrac{a \tau +b}{c \tau +d}, \tfrac{z}{c\tau + d} \right)= (c \tau + d)^k e^{\frac{2 \pi \ri c m  z^2}{c \tau + d}} \phi_{k,m}(\tau,z)   \,, \quad    \left(\begin{array}{cc}a&b\\ c& d \end{array}\right) \in {\rm SL}(2; \mathbb{Z})
	\label{mod}
	\ee
	under the action of the modular group, and are periodic in the elliptic parameter. Quasi-modularity under shifts of the elliptic parameters by $\tau$,
	\be
	\phi_{k,m}(\tau,z +\lambda \tau+ \mu)=e^{- 2 \pi \ri m (\lambda^2 \tau+ 2 \lambda z)}\phi_{k,m}(\tau,z) \quad \forall  \lambda, \mu \in \mathbb{Z}  \,,
	\label{shiftJacobi} 
	\ee
	follows from periodicity and \eqref{mod}. The two integers $k$ and $m$ are called the weight and the index of the Jacobi form, respectively.
	
	The ring $J$ of weak Jacobi forms of integer index (see \cite{EZ} for definitions) is generated by the two Jacobi forms $\phi_{-2,1}$ and $\phi_{0,1}$. These have the Taylor expansion
	\begin{eqnarray}
	\phi_{-2,1}(\tau,z) &=& -(2 \pi z)^2 + \frac{E_2 }{12}(2 \pi z)^4 + \frac{-5 E_2^2 + E_4}{1440} (2 \pi z)^6 + \frac{35 E_2^3 - 21 E_2 E_4 + 4 E_6}{362880}(2 \pi z)^8   \nn\\
	&&+ \mathcal{O}(z^{10}) \,, \nonumber \\ 
	\phi_{0,1}(\tau,z) &=& 12 - E_2 (2 \pi z)^2 + 
	\frac{E_2^2 + E_4}{24} (2 \pi z)^4 + \frac{-5 E_2^3 - 15 E_2 E_4 + 8 E_6}{4320} (2 \pi z)^6  + \mathcal{O}(z^8) \,. \nn \\
	\label{ABquasimodular}
	\end{eqnarray}    
	Note that the coefficients of this expansion take values in the ring of quasi-modular forms $\IC[E_2,E_4,E_6]$. Absence of the generator $E_2$ in the expansion coefficients implies that the Jacobi form is of index 0; examples of this type are necessarily meromorphic.

	The notion of Jacobi forms has been extended \cite{Wirthmuller:Jacobi,Bertola} by replacing the complex elliptic parameter $z \in \IC$ by an elliptic parameter $\boldsymbol{z}$ taking values in the complexified Cartan subalgebra $\mathfrak{h}_{\IC} \cong \IC^r$ of any rank $r$ simple Lie algebra $\mathfrak{g}$ other than $\mathfrak{e}_8$. The square in $z$ in the modular transformation \eqref{mod} is replaced by the norm of $\boldsymbol{z}$ induced by the (suitably normalized) Killing form, and quasi-modularity is imposed with regard to shifts $\lambda \tau + \mu$ with regard to elements $\lambda,\mu \in Q^{\vee}$ in the coroot lattice of $\mathfrak{g}$. As shown in \cite{Wirthmuller:Jacobi}, the space $J(\mathfrak{g})$ of such Jacobi forms which are invariant under the Weyl group action on $\boldsymbol{z}$ is again finitely generated. Generators were constructed explicitly in \cite{Bertola}.
	
	In \cite{DelZotto:2017mee}, it was shown in the case of $\mathfrak{g} = \mathfrak{a}_2, \mathfrak{d}_4$ that a freely generated subring of the ring of Weyl invariant Jacobi forms transforms well under diagram symmetries of the affine Lie algebra $\hat{\mathfrak{g}}$. Note that these transformations mix modular and elliptic parameters. This subring, which was dubbed $J^D(\mathfrak{g})$ in \cite{DelZotto:2017mee}, captures the fibral K\"ahler moduli dependence $\bm$ of the numerator of $\Zmodk$ in \eqref{eq:general_form}. The generators of $J^D(\mathfrak{a}_2)$ and $J^D(\mathfrak{d}_4)$ were determined explicitly in \cite{DelZotto:2017mee}. They have
	\be
	\mathrm{(weight, index)}_{\mathfrak{a}_2} = (0,3),(-2,3),(-6,6)
	\ee
	and
	\be
	\mathrm{(weight, index)}_{\mathfrak{d}_4} = (0,2),(-2,2),(-6,4),(-8,4),(-12,6)
	\ee
	respectively.
	
	The factors of exponentiated fibral K\"ahler parameters $Q_i$ which map $\Zmodk$ to $\Zk$, as introduced in \eqref{eq:enum_to_mod}, map $J^D(\mathfrak{g})$ to the ring $J^{\widehat{D}}(\mathfrak{g})$. The prefactors mapping generators of $J^D(\mathfrak{g})$ to generators of $J^{\widehat{D}}(\mathfrak{g})$, 
	\be \label{eq:map_D_to_Dhat}
	\begin{aligned}
		&J^D_{k,m}(\mathfrak{g}) & \longrightarrow \quad &J^{\widehat{D}}_{k,m}(\mathfrak{g}) 	 \,, \\ 
		&\phi_{k,m} & \longmapsto \quad&\Big(\prod_{i=1}^r Q_i^{a^{\vee}_i}\Big)^\frac{m}{2} \phi_{k,m}  \,,
	\end{aligned}
	\ee
	render these invariant under diagram symmetries, at the price of slightly perturbing their modular transformation behavior (in particular, $J^{\widehat{D}}(\mathfrak{g})$ is not a subring of $J(\mathfrak{g})$). 
	
	The case of $\mathfrak{g} = \mathfrak{e}_8$ is special, see \cite{Sakai:2011,Sakai:2017ihc}. For our purposes, we will only need to know that the ring of $\mathfrak{e}_8$ Weyl invariant Jacobi forms is generated by forms $A_i$, $i=1,\ldots, 5$, $B_i$, $i=1, \ldots, 6$ of index $i$. The generators $A_i$ and $B_i$ have weight 4 and 6, respectively. In the massless limit, we have $A_i \rightarrow E_4$, $B_i \rightarrow E_6$.

	\section{Determining the principal parts of $Z_3$ for the massless E-string}\label{sec: deg3}
  For simplicity, we only consider the massless limit of the E-string at base degree three. The generalization to the massive case is straightforward but  cumbersome. The decomposition \eqref{eq:rec-ref} of ${\cal Z}_{3}$ is given by
\be
{\cal Z}_{3} = {\cal F}_{3} + {\cal Z}_{1} {\cal Z}_{2} - \frac{1}{3} {\cal Z}_{1}^{3}\,,
\ee
where
\be
{\cal F}_{3} = \tf_{3}(\tau, z) + \frac{1}{3} \left(\frac{\eta^{36}(\tau) }{\eta^{12}(3\tau) } \right){\cal Z}_{1}(3\tau, 3z)\,.
\ee

We can read off the poles of ${\cal Z}_{3}$ from the expression \eqref{eq:denom_no_gauge} for the denominator specialized to $b_2(B)=1$, $k = 3$:
\be
z = \frac{1}{2}, \frac{\tau}{2}, \frac{\tau+1}{2}, \frac{m\tau + n}{3}\,,
\ee
where $m,n \in \{0, 1, 2\}$.

In order to determine the negative index Laurent data, assuming that we already know the expressions for ${\cal Z}_1$ and ${\cal Z}_2$, the only  input we need is the genus zero free energy at base-wrapping degree 3. This is given in \eqref{E3g0} for the massive case. The massless limit is taken by substituting $A_i \rightarrow E_4$, $B_i \rightarrow E_6$.

Following the strategy outlined in section \ref{sc:LC-unref}, we determine the principal parts of $\cZ_3$ as follows:

\paragraph{$\boldsymbol{z= 0}$} At the origin, all three terms in $Z_{3}$ contribute, yielding
\ba
\lc{0}{6}\,{\cal Z}_{3} &=& -\frac{\tpiI^{-6}}{6} E_{4}^{3}\,,\nn\\
\lc{0}{4}\,{\cal Z}_{3} &=& \frac{\tpiI^{-4}}{12} (E_{2}E_{4}^{3} + E_{4}^{2}E_{6})\,,\nn\\
\lc{0}{2}\,{\cal Z}_{3} &=& -\frac{\tpiI^{-2}}{155520} ( 3240 E_{2}^{2}E_{4}^{3} + 2359 E_{4}^{4} + 6480 E_{2}E_{4}^{2}E_{6} + 2645E_{4}E_{6}^{2} )\,.
\ea

\paragraph{$\boldsymbol{z= \frac{1}{3}, \frac{2}{3}}$} The second order poles at $z = \frac{1}{3}$, $z = \frac{1}{2}$ and $z = \frac{2}{3}$ appear due to the multi-covering contribution of ${\cal Z}_{1}$ in ${\cal F}_{3}$. Therefore, the Laurent coefficients can be simply read off as
\be
\lc{\frac{1}{3}}{2}\,{\cal Z}_{3} = \lc{\frac{2}{3}}{2}\,{\cal Z}_{3} = -\frac{\tpiI^{-2}}{27} \frac{\eta^{36}(\tau)}{\eta^{12}(3\tau)} E_{4}(3\tau)\,.\\
\ee

\paragraph{$\boldsymbol{z= \frac{1}{2}, \frac{\tau}{2}, \frac{\tau + 1}{2}}$} These poles are due to $\cF_3$ and $\cZ_1 \cZ_2$. Relating them via modularity to the Laurent coefficients at the real poles, we obtain
\ba
\lc{\frac{1}{2}}{2}\,{\cal Z}_{3} &=& \frac{\tpiI^{-2}}{8} \left(\frac{\eta^{24}(\tau)}{ \eta^{12}(2\tau)}\right) \frac{E_{4}(\tau)E_{4}(2\tau)}{\phi_{-2,1}(\tau, \frac{1}{2})}\,, \nn\\
\lc{\frac{\tau}{2}}{2}\,{\cal Z}_{3} &=& -\frac{\tpiI^{-2}}{2}q^{5/4} \left(\frac{\eta^{24}(\tau)}{\eta^{12}(\frac{\tau}{2})}\right) \frac{E_{4}(\tau)E_{4}(\frac{\tau}{2})}{\phi_{-2,1}(\tau,\frac{\tau}{2})}\,,\nn\\
\lc{\frac{\tau}{2}}{1}\,{\cal Z}_{3} &=& -\frac{1}{2\pi \ri} q^{5/4} \left(\frac{\eta^{24}(\tau)}{\eta^{12}(\frac{\tau}{2})}\right) \frac{[3\phi_{-2,1}(\tau, \frac{\tau}{2}) - \partial_{z}\phi_{-2,1}(\tau, \frac{\tau}{2})]E_{4}(\tau)E_{4}(\frac{\tau}{2})}{\phi_{-2,1}(\tau, \frac{\tau}{2})^{2}}\,,\nn\\
\lc{\frac{\tau + 1}{2}}{2}\,{\cal Z}_{3} &=& -\ri\frac{\tpiI^{-2}}{2} q^{5/4} \left(\frac{\eta^{24}(\tau) }{\eta^{12}(\frac{\tau + 1}{2})}\right) \frac{E_{4}(\tau) E_{4}(\frac{\tau + 1}{2})}{\phi_{-2,1}(\tau,\frac{\tau + 1}{2})}\,,\nn\\
\lc{\frac{\tau + 1}{2}}{1}\,{\cal Z}_{3} &=& -\frac{1}{2\pi} q^{5/4} \left(\frac{\eta^{24}(\tau) }{\eta^{12}(\frac{\tau + 1}{2})}\right) \frac{[3\phi_{-2,1}(\tau, \frac{\tau + 1}{2}) - \partial_{z}\phi_{-2,1}(\tau, \frac{\tau + 1}{2})]E_{4}(\tau)E_{4}(\frac{\tau + 1}{2})}{\phi_{-2,1}(\tau, \frac{\tau + 1}{2})^{2}}\,.\nn\\
\ea

\paragraph{$\boldsymbol{z = \frac{m\tau + n}{3} \,, (m \neq 0)}$} The remaining six poles are due to $\tf_{3}$ only. We obtain
\ba
\lc{\frac{\tau}{3}}{2}\,{\cal Z}_{3}  &=& \frac{1}{q^{2}}\lc{\frac{2\tau}{3}}{2}\,{\cal Z}_{3} = -\frac{\tpiI^{-2}}{3}q^{2/3} \left(\frac{\eta^{36}(\tau)}{\eta^{12}(\frac{\tau}{3})}\right) E_{4}(\tfrac{\tau}{3})\,,\nn\\
\lc{\frac{\tau}{3}}{1}\,{\cal Z}_{3}  &=& \frac{1}{2q^{2}}\lc{\frac{2\tau}{3}}{1}\,{\cal Z}_{3} = -4\frac{\tpiI^{-1}}{3}q^{2/3} \left(\frac{\eta^{36}(\tau)}{\eta^{12}(\frac{\tau}{3})}\right) E_{4}(\tfrac{\tau}{3})\,,\nn\\
\lc{\frac{\tau + 1}{3}}{2}\,{\cal Z}_{3}   &=& \frac{1}{q^{2}}\lc{\tfrac{2\tau + 2}{3}}{2}\,{\cal Z}_{3} = -\frac{\tpiI^{-2}e^{\pi i/3} }{3}q^{2/3} \left(\frac{\eta^{36}(\tau) }{\eta^{12}(\frac{\tau + 1}{3})}\right) E_{4}(\tfrac{\tau + 1}{3})\,,\nn\\
\lc{\frac{\tau + 1}{3}}{1}\,{\cal Z}_{3}  &=& \frac{1}{2 q^{2}}\lc{\frac{2\tau + 2}{3}}{1}\,{\cal Z}_{3} = -4\frac{\tpiI^{-1}e^{\pi i/3}}{3}q^{2/3} \left(\frac{\eta^{36}(\tau)}{\eta^{12}(\frac{\tau + 1}{3})}\right) E_{4}(\tfrac{\tau + 1}{3})\,,\nn\\
\lc{\frac{\tau + 2}{3}}{2}\,{\cal Z}_{3}  &=& \frac{1}{q^{2}}\lc{\frac{2\tau + 1}{3}}{2}\,{\cal Z}_{3} = -\frac{\tpiI^{-2}e^{2\pi i /3} }{3}q^{2/3} \left(\frac{\eta^{36}(\tau)}{\eta^{12}(\frac{\tau + 2}{3})}\right) E_{4}(\tfrac{\tau + 2}{3})\,,\nn\\
\lc{\frac{\tau + 2}{3}}{1}\,{\cal Z}_{3}  &=& \frac{1}{2 q^{2}}\lc{\frac{2\tau + 1}{3}}{1}\,{\cal Z}_{3} = -4\frac{\tpiI^{-1}e^{2\pi i /3} }{3}q^{2/3} \left(\frac{\eta^{36}(\tau)}{\eta^{12}(\frac{\tau + 2}{3})}\right) E_{4}(\tfrac{\tau + 2}{3})\,.
\ea
We have checked these results by using them to compute the full Jacobi form $\cZ_3$, following the discussion of section \ref{s:explicit_reconstruction}, and checking against the results in \cite{Gu:2017ccq}.

\section{Index bilinear of the massive higher rank E-strings} \label{s:index_bilinear}

The index bilinear for the elliptic genus of the 
massive higher rank E-strings at base degree $\bk$ is
\begin{align}
  M_{\bk}(z_L,z_R,\bm)
  &= -\frac{1}{2}\(-\sum_{IJ}C_{IJ}k_I k_J +
    \sum_{I}(2-n_I)k_I\)z_L^2 \nn\\
  &\phantom{=}- \frac{1}{2}\(\sum_{IJ} C_{IJ}k_I k_J +
    \sum_{I}(4-n_I)k_I\)z_R^2 \nn\\
  &\phantom{=}+  \sum_i k_i \,m^2 +  \frac{k_0}{2}   (\bm,\bm)_{\mathfrak{e}_8}\ ,
\end{align}
where $C_{IJ}$, is the intersection matrix of the base curves,
$-n_I$ their self-intersection numbers. We have indexed the base curves yielding M-strings by the label $I=i$, and the curve on which the chain terminates to yield the E-string by $I=0$. We have furthermore denoted the flavor fugacities associated to the chain of M-strings by $m$, and the $\mathfrak{e}_8$ flavor fugacity associated to the E-string by $\bm$. We see that both the index of the flavor fugacities and the sum of the indices of
$z_L$, $z_R$ are linear in $\bk$.

\section{The denominator of $\Zmodk$ in terms of generators of $J^D(\mathfrak{g})$ for $\mathfrak{g}=\mathfrak{a}_2$ up to $k=2$}
The results in section \ref{ss:transf_unref} for the topological free energy on the elliptic fibration over $\cO(-3) \rightarrow \IP^1$ (the geometry which engineers the minimal 6d SCFT with gauge group $SU(3)$ in F-theory) rely on expressing the denominator \eqref{Z_jacobi_structure-2} in terms of the generators of the ring $J^D(\mathfrak{a}_2)$. As the ring is finitely generated, this is an exercise in Taylor expanding the denominator to appropriately high order to fix the finite number of expansion coefficients in these generators. We here gather the results up to $k=2$.

\paragraph{k=1} $\alpha_1(0) =1$

\be
\prod_{\alpha \in \Delta_+} \phi_{-2,1}(\tau, m_{\alpha}) = - \phi_{-6,6} \,.
\ee

\paragraph{k=2} $\alpha_2(0) =1$, $\alpha_2(1) =1$

\be
\begin{aligned}
	\prod_{\alpha \in \Delta_+}  \big( \phi_{-2,1}&(\tau, z + m_\alpha) \phi_{-2,1}(\tau, z - m_\alpha) \big) = \frac{A^6 \phi_{0,3}^4}{1296}+\frac{A^5 B \phi_{0,3}^3 \phi_{-2,3}}{5184}+\frac{A^4 B^2 \phi_{0,3}^2 \phi_{-2,3}^2}{55296} \\
	&+\frac{A^3 B^3
		\phi_{0,3} \phi_{-2,3}^3}{1327104}+\frac{A^2 B^4 \phi_{-2,3}^4}{84934656}+\frac{A^3 B^3 \phi_{0,3}^2 \phi_{-6,6}}{31104}-\frac{A^5 B E_4
		\phi_{0,3}^2 \phi_{-6,6}}{10368} \\
	&+\frac{A^6 \ E_6 \phi_{0,3}^2 \phi_{-6,6}}{15552}+\frac{A^2 B^4 \phi_{0,3} \phi_{-2,3} \phi_{-6,6}}{248832}-\frac{A^4 B^2 E_4 \phi_{0,3} \phi_{-2,3} \phi_{-6,6}}{82944} \\ 
	&+\frac{A^5 B E_6 \phi_{0,3} \phi_{-2,3} \phi_{-6,6}}{124416}+\frac{\ A B^5 \phi_{-2,3}^2
		\phi_{-6,6}}{7962624}-\frac{A^3 B^3 E_4 \phi_{-2,3}^2 \phi_{-6,6}}{2654208} \\
	&+\frac{A^4 B^2 E_6 \phi_{-2,3}^2 \phi_{-6,6}}{3981312} +\frac{B^6 \phi_{-6,6}^2}{2985984}-\frac{A^2 B^4 E_4 \phi_{-6,6}^2}{497664}+\frac{A^4 B^2 E_4^2 \phi_{-6,6}^2}{331776} \\ 
	&+\frac{A^3 B^3 E_6
		\phi_{-6,6}^2}{746496}-\frac{A^5 B E_4 E_6 \phi_{-6,6}^2}{248832}+\frac{A^6 E_6^2 \phi_{-6,6}^2}{746496} \,.
\end{aligned}
\ee
We have here introduced the notation $A=\phi_{-2,1}(\tau,z)$, $B= \phi_{0,1}(\tau,z)$. The arguments of $\phi_{0,3}, \phi_{-2,3}, \phi_{-6,6}$ are of course $(\tau, \bm)$.

	\bibliography{E_from_poles}
	
\end{document}